\documentclass[letterpaper]{article}

\usepackage[letterpaper,top=3cm,bottom=2cm,left=3cm,right=3cm,marginparwidth=1.75cm]{geometry}

\usepackage{amsmath,amsthm,amssymb,amsfonts}
\usepackage{graphicx}
\usepackage[colorinlistoftodos]{todonotes}
\usepackage[colorlinks=true, allcolors=blue]{hyperref}

\usepackage{amsmath}
\usepackage{amssymb}
\usepackage{graphics,graphicx}
\usepackage{epstopdf}
\usepackage{lineno,hyperref}
\usepackage{color}
\usepackage{caption,bm}
\usepackage{subcaption}
\usepackage[normalem]{ulem}
\usepackage{pifont}
\usepackage{booktabs,tabularx}

\newcommand{\req}[1]{(\ref{#1})}

\newcommand{\fref}[1]{Fig.~\ref{#1}}

\newcommand{\stkout}[1]{\ifmmode\text{\sout{\ensuremath{#1}}}\else\sout{#1}\fi}

\title{Fate of a bulge in an inflated hyperleastic tube: theory and experiment}

\author{Masoud Hejazi\footnote{Department of Mechanical engineering, University of British Columbia, Vancouver, BC V6T 1Z4, Canada} \\
	York Hsiang\footnote{Department of Surgery, University of British Columbia, Vancouver, BC V6T 1Z4, Canada}\\
A. Srikantha Phani\footnote{Corresponding Author: Department of Mechanical engineering, University of British Columbia, Vancouver, BC V6T 1Z4, Canada. phani@mech.ubc.ca . }\\
}

\date{September 2020}
\begin{document}
	\maketitle
	
	%\begin{abstract}
	%Your abstract.
	%\end{abstract}
	\section*{{Abstract}}
	Mechanical instability in a pre-tensioned finite hyperelastic tube subjected to a slowly increasing internal pressure produces a spatially localized bulge at a critical pressure. The subsequent fate of the bulge, under continued inflation, is critically governed by the end-conditions, and the initial tension in the tube. In a tube with one end fixed and a dead weight attached to the other freely moving end, the bulge propagates axially at low initial tension, growing in length and the tube relaxes by extension. Rupture occurs when the tension is high. In contrast, the bulge formed in a tube, initially stretched and held fixed at both its ends can buckle or rupture, depending on the amount of initial tension. Experiments on inflated latex rubber tubes are presented for different initial tensions and boundary conditions. Failure maps in the stretch parameter space and in stretch-tension space are constructed, by extending the theories for bulge formation and buckling analyses to the  experimentally relevant boundary conditions. The fate of the bulge according to the failure maps deduced from the theory is verified; the underlying assumptions are critically assessed. It is concluded that buckling provides an alternate route to relieve the stress built up during inflation. Hence buckling, when it occurs, is a protective fail safe mechanism against the rupture of a bulge in an inflated elastic tube. 
	\section*{keyword}
	Localization, Stability, Bulge, Buckling, Aneurysms
	
	\section{Introduction}
	\label{introduction}
\begin{figure}[!h]
		\center
		\includegraphics[width=0.8\textwidth]{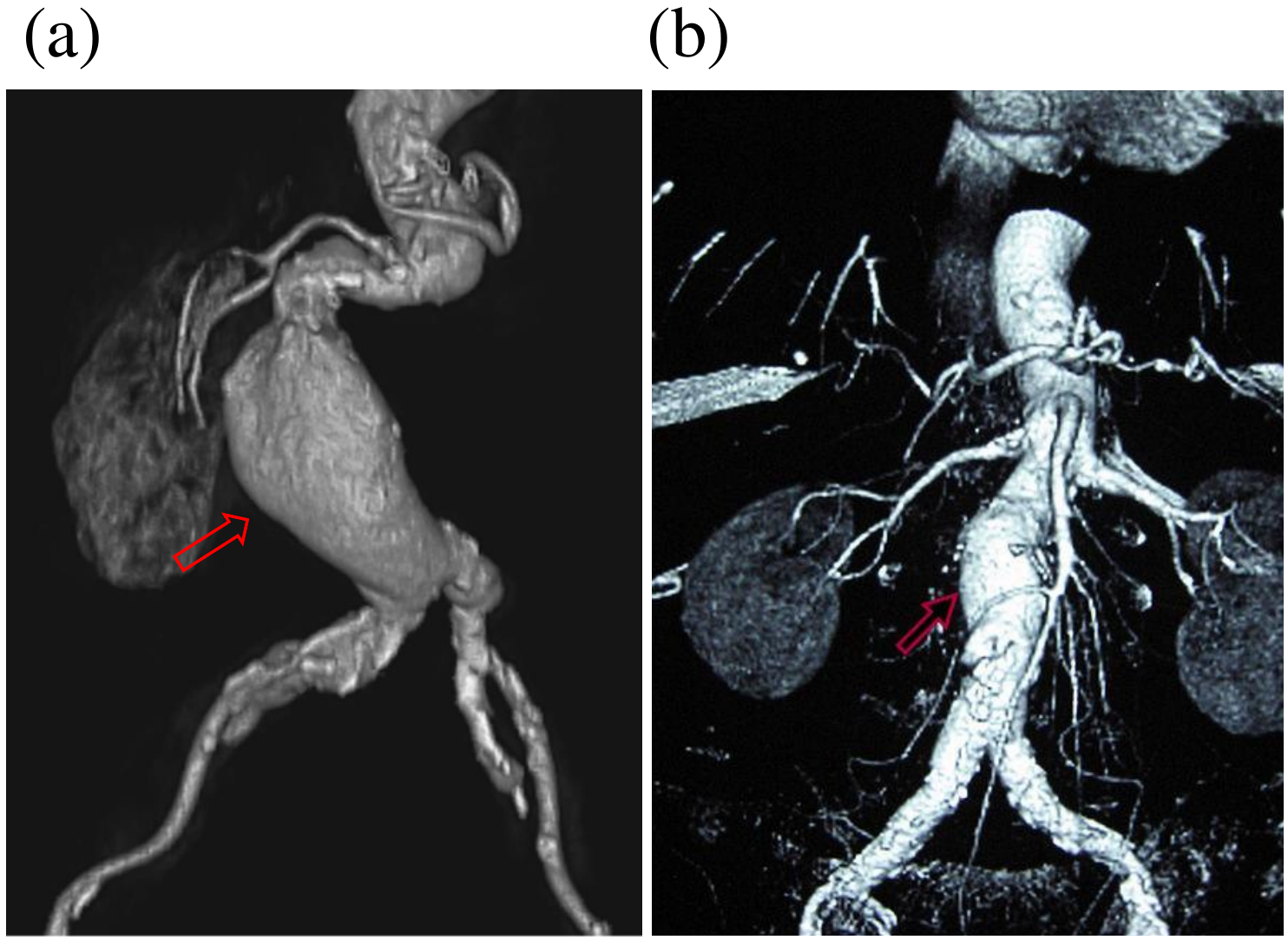}
		\caption{(a) Buckled aneurysm, open source image from: epos.myesr.org/. (b) Unbuckled aneurrysm, open source image from: wikimedia.org.}
		\label{int_aneu}
	\end{figure}

	The objective of this work is to understand the fate of a spatially localized bulge, initiated at a critical internal pressure in a pre-stretched finite prismatic cylindrical tube made of a hyperelastic material. The bulge can propagate axially along the tube, grow radially in size and rupture, or buckle laterally  in a fail safe manner. A combination of these phenomena can also occur. There is considerable literature on the initiation and propagation of bulges in pressure vessels~\cite{Kyriakides2007,chater1984}. Despite their apparent simplicity, inflated balloons and tubes have produced surprisingly relevant experimental observations and illustrations for thermodynamic paradigms~\cite{Muller2004}. Interest in understanding the growth and rupture of aneurysms in human blood vessels motivated this study. Here, a local weakening of an arterial wall is presumed to initiate an axially localized bulge, called an aneurysm. In clinical practice, some aneurysms are observed in a laterally buckled state.  Evidence suggests that such aneurysms do not rupture despite their size exceeding the minimum recommended size for surgical interventions~\cite{Fillinger2004} (see \fref{int_aneu}). The material properties of the  arteries are complex and often difficult to measure and model \emph{in vivo}, given the complexity of human vascular anatomy and {anisotropic fiber reinforced structure of the blood vessel}~\cite{Holzapfel:2019}. Pathogenesis of the aneurysm leads to collagen fiber degradation, reduces anisotropy and diminishes the exponential stiffening response under mechanical stress \cite{sakalihasan2018,kobielarz2015}. Hence, An alternate path to qualitatively understand the phenomenon is to replicate  buckled aneurysms under controlled settings in a laboratory. Such  experiments on an inflated latex rubber tube ( {representing the arterial aneurysm with hyperplastic properties}) under an imposed tension are reported here, albeit within an isotropic elasticity setting of large deformations in the present study. The post bulge response in these tests is fracture or buckling, depending on the magnitude of pre-tension and boundary conditions.  Outcomes in these controlled experiments are compared against the predictions based on mechanics models, incorporating diffuse interface modeling (DIM)~\cite{lest2018,lest2020} and bifurcation analysis~\cite{shield1972,haughton1979a,haughton1979b,wang2019,ye2020}. Portraits spanned by axial and circumferential stretch axes and the axial tension and circumferential strain axes emerge from the present analyses, demarcating the dangerous rupture and fail safe buckling regimes.
	
	Pressurized cylindrical vessels  are found in many  technological  applications: in oil and gas pipes~\cite{palmer1975, Kyriakides2007}, nuclear plants; and also in biology in blood vessels, for example. The stability of pressure vessels under combined loading is an active area of research despite its long history~\cite{chater1984,kyr1991,wang2019,HE2020}, not all of which can be surveyed here as  the literature spans at least  four decades.  Pertinent results are recalled here.  That an internally pressurized India rubber tube exhibits a local bulge  has been experimentally demonstrated by Mallock in 1891, who also reported an initial analysis~\cite{mallock1891}.  Bent submarine pipelines also show a localized kink buckling, which can transform and propagate  when driven by an external pressure~\cite{palmer1975,Calladine1988,Kyriakides1993}.  The bulge problem  in a hyperelastic tube  has received wide  attention since its origins in offshore pipelines.  
	Inflating a cylindrical hyperelastic tube under prescribed volume starts first with a uniform expansion regime, followed by the formation of a long wavelength bulge along the entire length of the tube~\cite{chater1984,Kyriakides1993} which localizes at a critical  (maximum) internal pressure.  In the post localization regime, the bulge propagates radially and axially with an associated propagation pressure, below the bulge initiation pressure. This propagation pressure  can be estimated using Maxwell's equal-area rule~\cite{chater1984}. Ideally, the propagating bulge invades the entire tube length. In reality, however, the tube can rupture or buckle depending on the boundary conditions at the ends and the initial tension. Experiments on bulge  propagation in latex tubes have been conducted under a dead weight attached to one end and
	the fluid (air) injected at the other end~\cite{kyr1990,kyr1991}. They showed the effect of the axial tension on the shape of the bulge profile, maximum pressure, and the circumferential stretch at the onset of bifurcation  and used a nonlinear-membrane model~\cite{shield1972,haughton1979a,haughton1979b} to corroborate experiments. The analysis of inflation of a cylindrical tube  under an axial load~\cite{haughton1979a,haughton1979b} suggests that for a fixed-free case, the length of the tube monotonically increases during inflation. For a fixed-fixed case, however, the pre-stretched length of the  tube is held constant, and the axial force may increase, decrease, or remain constant depending on the pre-stretched length of the tube, or equivalently pre-imposed axial tension. In the membrane approximation, the influence of bending stiffness on the bifurcation is negligible except for a very thick tube.  Membrane assumption has been used in~\cite{lest2018} to construct a diffuse interface model (DIM) to address large deformations, and it is asymptotically exact for tubes with high length to diameter ratio. The application of this method to study  bulge formation in tubes with fixed-free end conditions is pursued in~\cite{lest2018,lest2020}.  In an extensive study by Fu and colleagues, the bifurcation analysis has been revisited by developing a weakly non-linear model to study the post-bifurcation response of infinite tubes~\cite{fu2008,fu2010,pearce2010,fu2012}. Through these studies, imperfection sensitivity and loading (volume/pressure control) have been investigated. Following these works,  a form of bifurcation condition based on the Jacobian function of the internal pressure and axial force, which in turn are expressed in terms of the axial and circumferential stretches, is obtained~\cite{fu2016}. {Using this method}, the effect of multilayering~\cite{liu2019} and fiber-reinforcement~\cite{wang2018} has been examined. In all of these studies, the footprint of boundary conditions is evident. For instance, it is shown by analysis~\cite{wang2018} that the localized bulge formation with fixed-free boundary conditions is not possible for fiber reinforced tubes. Furthermore, the bifurcation conditions associated with the bulge  have been verified by experiment for both fixed-fixed and fixed-free boundary conditions~\cite{wang2019}. A comparison between the weakly non-linear model and membrane theory can be found in~\cite{ye2020}.  Finally, in passing, one should mention the studies on the bulge formation in dielectric tubes, which use a variational method based  on the thermodynamic free energy of the system~\cite{lu2015}, see also~\cite{dorfmann2019} for an extensive discussion on experimental and analytical methods.
	
	In {most} the experiments reported thus far, the lateral buckling does not occur, since the end weight producing tension is free to move to accommodate unloading in~\cite{kyr1990,kyr1991}, for example. It is possible, both in applications such as an aneurysm or supported pipes, to observe a different boundary condition where both ends are held fixed. This can be accomplished,  for example, by holding the tube between spatially fixed jaws of a material testing system  to impose a predefined axial stretch. {Although some studies incorporated the fixed-fixed boundary conditions and observed post-bulge buckling in limited number cases \cite{wang2019,gonccalves2008}}. Very little seems to be known about the post-bulge buckle propagation and formation mechanism in this restrained inflation case, which is the subject of this article.  We present the  experiments first in section 2, followed by  an analysis in section 3, a discussion comparing model and measurements in section 4, and end with concluding remarks in section 5.

\section{Experiments}
A series of quasi-static inflation tests on  latex rubber tubes will now be presented  to investigate the effect of boundary conditions on the bulge profile and post-bulge response.  First, quasi-static uniaxial tests with different tube lengths  are performed to identify the constitutive constants ($\alpha_i$) in the Ogden's strain energy density function (SEF)
\begin{equation}
{W}=\displaystyle\sum_{i=1}^{3}\frac{s_i}{\alpha_{i}}(\lambda_{1}^{\alpha_i}+\lambda_{2}^{\alpha_i}+\lambda_{3}^{\alpha_i}-3), \label{osef}
\end{equation}
where $W$ is the strain energy function, and  $\lambda_j$ are the principal stretches.  The measured stress-stretch curves are fitted following the established procedures  {to calibrate the SEF parameters based on the uniaxial tensile test~ }\cite{ogden2004}. See Appendix A for details. The geometric and material properties of the samples tested are as given in Table.~\ref{T:1}. 
\begin{table}[!h]
	\caption{Geometric and material properties of the rubber tubes studied. The constitutive  constants $s_i$ and $\lambda_i$ are Ogden's SEF parameters arising in~\req{osef}.}\label{T:1}
	\vspace{-3mm}
	\begin{tabular}{p{7cm}p{7cm}}
		\toprule
		Internal diameter, $D$& 6.5 $mm$\\
		Length, $2L$& 20 $cm$\\ %10--20 $cm$
		Tube wall thickness, $H$ &  1.5 $mm$\\
		Ultimate uniaxial stretch, $\lambda_{ut}$& $7.85\pm0.8$ \\
		Ogden's SEF parameters& $\alpha_{1}=1.04$; $\alpha_{2}=1.33$; $\alpha_{3}=4.46$ \\
		&   $s_{1}=704.8$; $s_{2}=373.2$; $s_{3}=2.8$ ($s_i$ in kPa) \\
		\bottomrule
	\end{tabular}
\end{table}
Note that the wall thickness to diameter ratio is small ( $H/D<0.23$) so that the membrane approximation for bulge analysis~is justified according to~\cite{fu2016} .

\begin{figure}[!h]
	\center
	\includegraphics[width=0.8\textwidth]{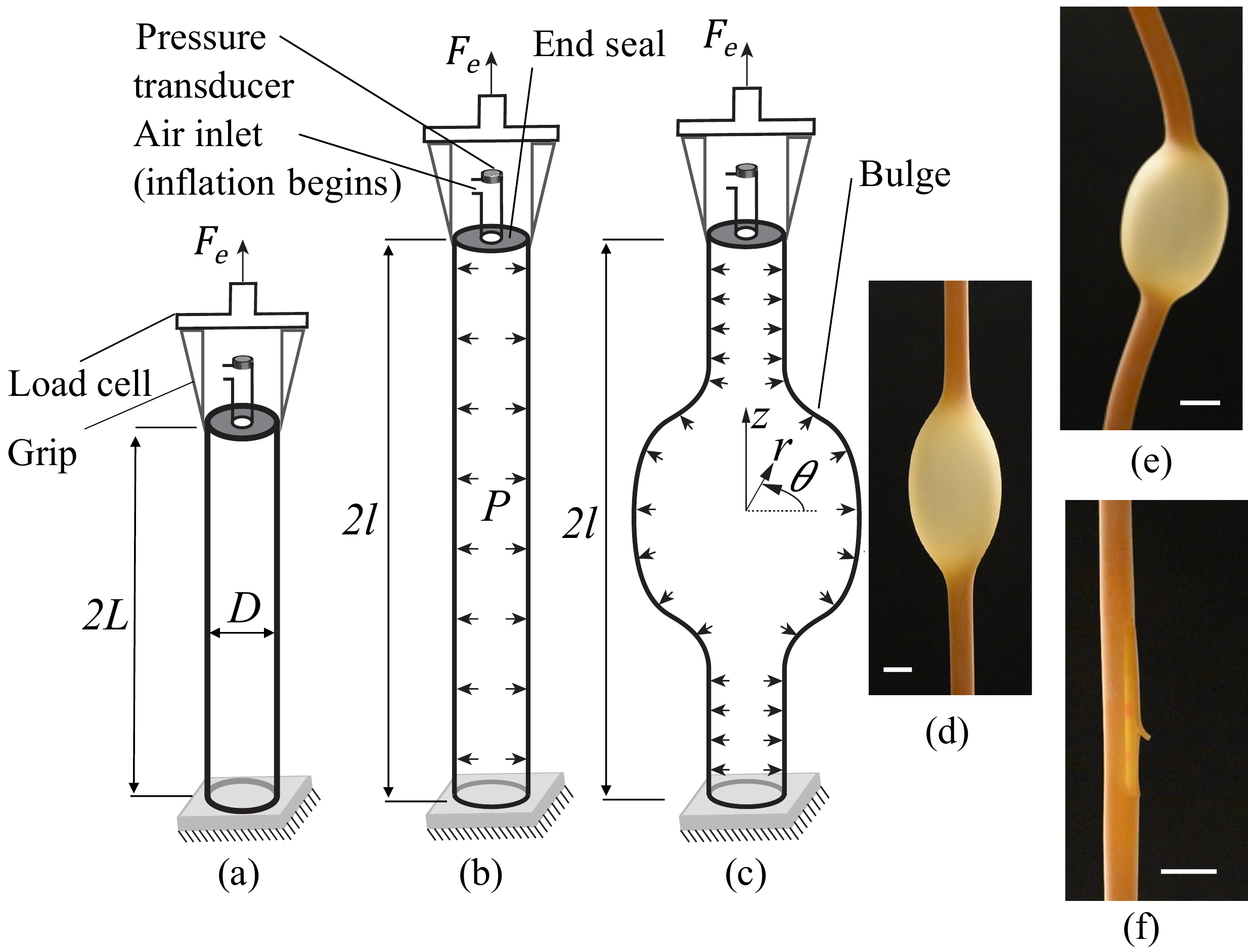}
	\caption{Bulging and rupture of a pre-stretched hyper-elastic rubber tube under gradual pneumatic inflation  in a typical displacement control experiment. The experimental sequence is:  (a) a pre-stretch is applied first and held constant throughout the experiment; (b) The stretched tube is gradually inflated by increasing the internal air pressure ($P$); (c) Bulge forms at a critical pressure for a given stretch; (d) Experimentally observed bulge which can either buckle or rupture, depending on the initial stretch;  (e) Further inflation leads to buckling for low values of pre-stretch; (f) Further inflation  leads to rupture for higher values of pre-stretch. The axial force ($F_e$) is measured in the tensile testing machine (INSTRON, model 5969) and the pressure $P$ is measured using a transducer near the moving (upper) grip of the tensile tester.  Scale bars in (d), (e) and (f) are 1 cm. (Color online)}
	\label{Fig:1}
\end{figure}

\begin{figure}[!h]
	\center
	\includegraphics[width=0.8\textwidth]{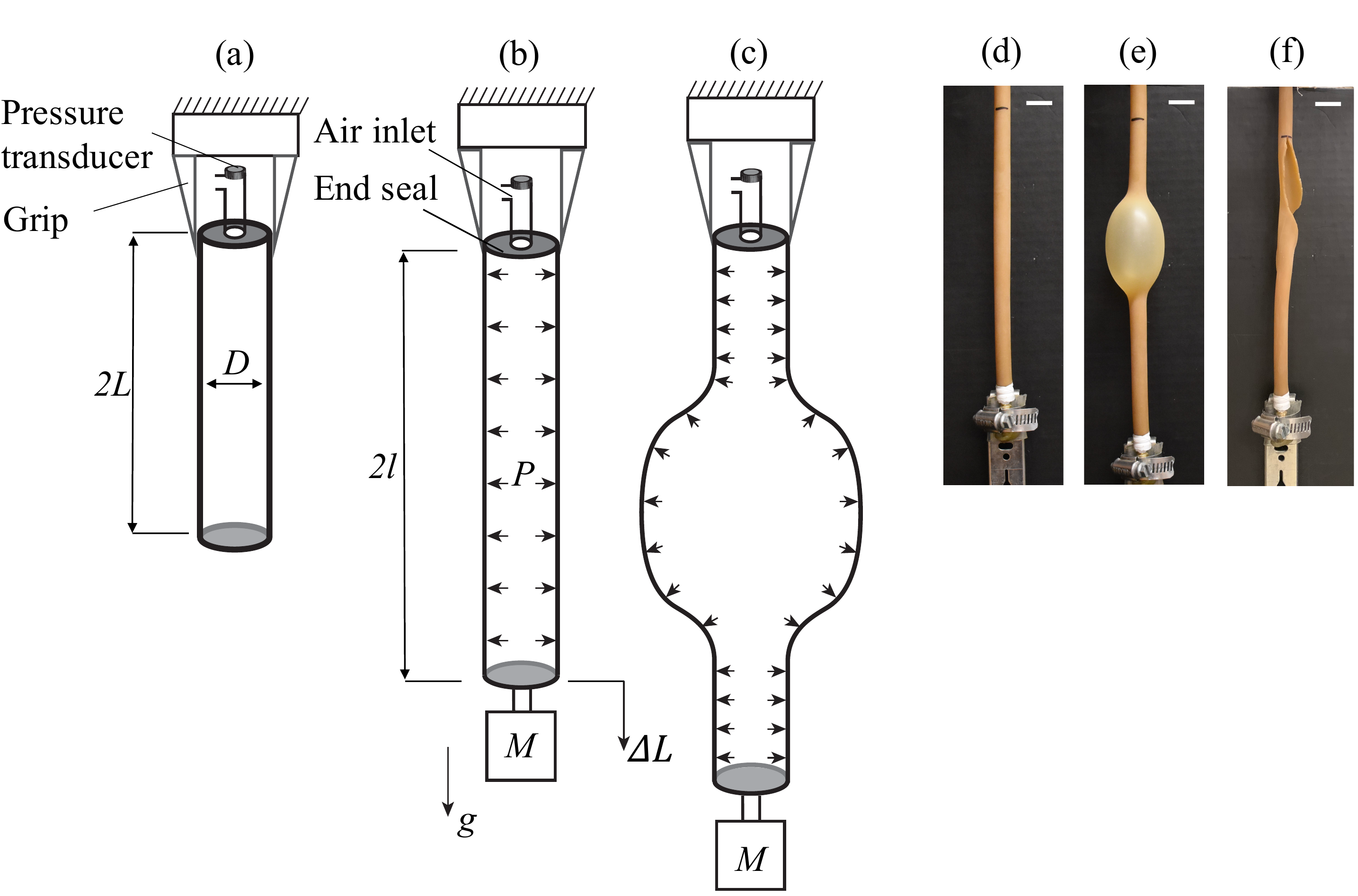}
	\caption{Bulging and rupture of a pre-stretched hyper-elastic rubber tube under gradual pneumatic inflation  in a typical force control experiment using a dead weight. The experimental sequence is: (a) Initial  unstretched configuration with the upper end fixed; (b) Mass  ($M$) is attached to maintain a constant axial gravitational force; (c) Slow inflation of the tube leads to the formation of a localized bulge for a given $M$; (d) Observed bulge in a typical experiment; (e) Further inflation leads to rupture without buckling. Scale bars in (d) and (e)  are 1 cm.}
	\label{Fig:2}
\end{figure}

The bulge experiments fall into two groups with respect to  axial loading and boundary conditions: (1)  displacement control experiments with a fixed-fixed boundary condition in a material testing system; (2) force control experiments under a fixed-free boundary condition with a dead load attached to the free end. The displacement control set up is shown in~\fref{Fig:1}, where the upper and the lower ends represent the grips of a material testing system. The experiment proceeds in the following stages as shown in~\fref{Fig:1}: (a) a pre-stretch is applied to achieve the desired initial axial tension; (b) the inflation starts from the port at the top air inlet, while the pre-stretch is maintained by the machine; (c) the localized bulge forms and propagates; (d) the tube buckles (e) or ruptures (f), depending on the amount of initial tension. Similarly, the force control set up is shown in~\fref{Fig:2},  and it has four stages: (a) a dead weight is attached to the tube to maintain the desired axial tension; (b) the inflation starts; (c) the localized bulge forms and propagates; (d) the tube ruptures. The main difference between the two categories is the boundary conditions (BCs). In the displacement control (fixed-fixed BCs), the stretched length of the tube is held constant and the axial tension varies during the experiment. However, in the force control (fixed-free BCs) the axial tension is held constant by the dead weight and the length of the tube can change as one of the ends is attached to a slider and  hence free to move vertically. A digital camera (Nikon DSLR D7200) camera records the deformation of the tube during injection and a pressure transducer records the internal pressure.  An edge detection algorithm and digital image correlation (DIC) in the MATLAB$^\text{\textregistered}$~\cite{blaber2015ncorr} image processing toolbox is used for calculating the circumferential and  the axial stretches,  the internal volume based on incompressibility assumption, and  the bulge profile for further analysis. For  the displacement control experiments, an Instron tensile tester (Model 5965) applies a prescribed pre-stretch by adjusting the relative positions of the grips, which are subsequently held fixed so that the final length is constant. A load cell measures the variable axial tension during inflation.

\section{Analyses} \label{s:analysis}
This section considers the analysis of bulging, buckling and rupture. In the bulging analysis, the walls of the inflated cylindrical  vessel  are idealized as  membranes with negligible bending stiffness, given the low wall thickness to diameter ratio.  The buckling analysis takes bending stiffness, however small, into account. Rupture analysis uses the strain energy density function based failure criterion. Two modeling frameworks are taken up to study bulging.  The first, axisymmetric membrane theory relies on a set of four coupled ordinary differential equations (ODE). The second method, based on the diffuse interface model (DIM), replaces the four ODEs  by a single scalar master equation. The bifurcation point can be evaluated efficiently using DIM, while the bulge profile of a finite tube  given by the axisymmetric membrane theory is more realistic. Both methods will be modified in the implementation of the boundary conditions.

\subsection{Axisymmetric membrane theory} 
\begin{figure}[!ht]
	\centering
	\includegraphics[width=0.6\textwidth]{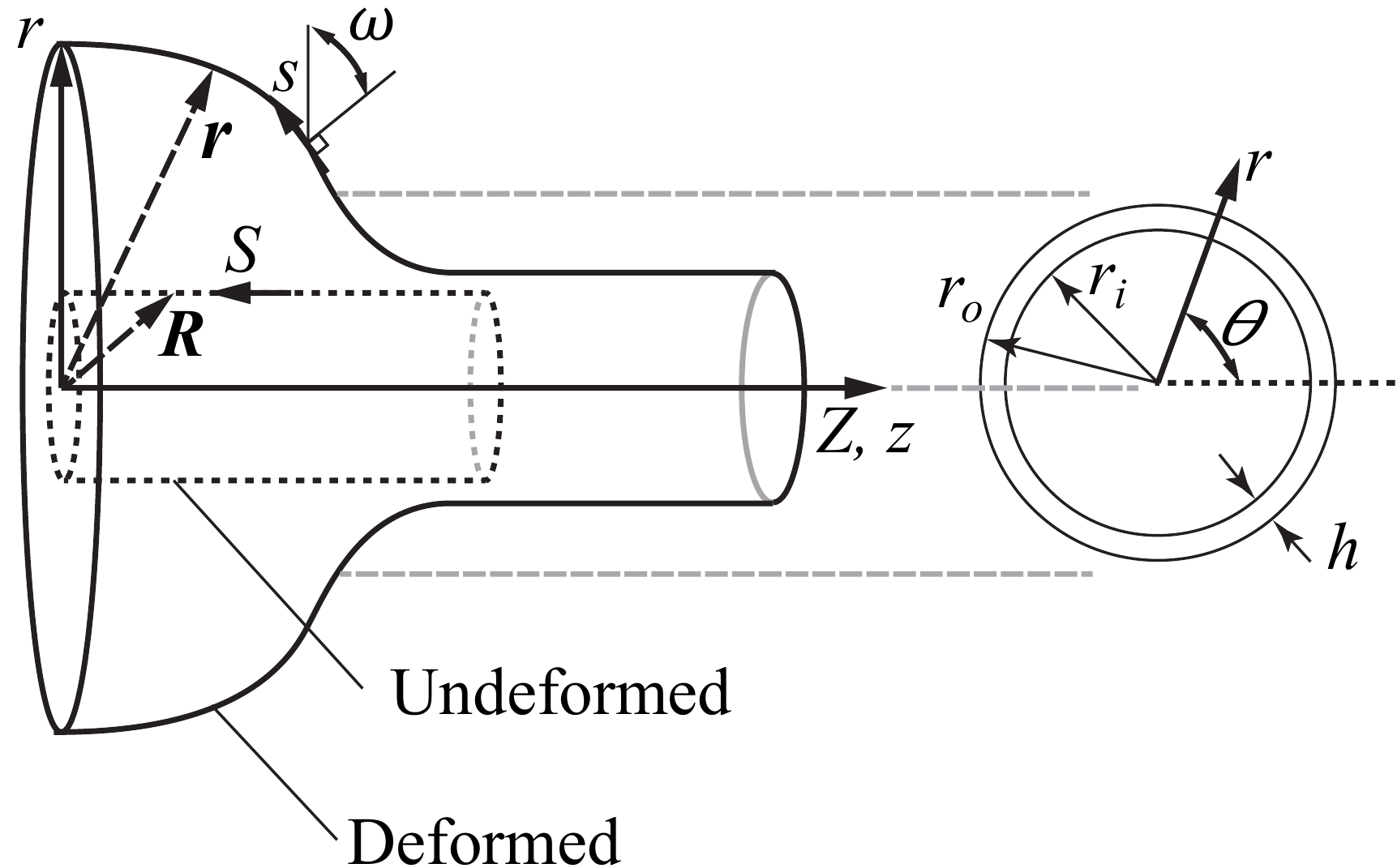}
	\caption{Deformed and undeformed configurations of a finite deformation in an axisymmetric membrane element sketched for a half of the tube.  The total length of the tube is $2L$, and the region $0\le Z \le L$ is shown with the origin located at the crown of the bulge. The meridional coordinate ($S$ in undeformed and $s$ in deformed) and radius $R$ can be  function of $Z$, providing the ability to introduce imperfection. The principal curvatures are $\kappa_{1}=-{d\omega}/{ds}$ and $\kappa_{2}={\cos{\omega}}/{r}$.}
	\label{Fig_1}
\end{figure}

Adopting a cylindrical coordinate systems shown in~\fref{Fig_1},  the triad $\bm{R}=(R,\Theta,Z)$ locates a material point in the undeformed configuration, and the triad $\bm{r}=(r,\theta=\Theta,z)$ the same point in the deformed configuration. Making use of symmetry it is sufficient to analyze the region $0\le z \le L$ as sketched in~\fref{Fig_1}, where the origin of the co-ordinate system lies at the crown of the bulge.

Explicitly accounting for the incompressibility, the  thickness averaged principal stretches of the axisymmetric membrane   are $(\lambda_{1},\lambda_{2},\lambda_{3})=(\lambda_{z}, \lambda_{\theta}, ({\lambda_{z}\lambda_{\theta}})^{-1})$, where the associated thickness averaged stretches are  $\lambda_{z}={ds}/{dS}$ and $\lambda_{\theta}={r}/{R}$. Note that $s$ and $S$ are respectively along the meridonial arc length of the deformed, and the undeformed tube  as shown in~\fref{Fig_1}. Hence forward, the subscripts $1$ and $2$ are used interchangeably with $z$ and $\theta$, respectively. The constitutive strain energy density function in the reference configuration and the Cauchy stresses follow as 
\begin{equation}
{W}(\lambda_{z},\lambda_{\theta},\lambda_r)={W}(\lambda_{z},\lambda_{\theta},\displaystyle\frac{1}{\lambda_{z}\lambda_{\theta}})\equiv \hat{W}(\lambda_{z},\lambda_{\theta}), \quad \sigma_{i}=\lambda_{i}\frac{\partial{\hat{W}}}{\partial \lambda_{i}}=\lambda_i\hat{W}_{\lambda_i}, ~i=1,~2,
\end{equation}
where, $\sigma_{1}=\sigma_{z}$, $\sigma_{2}=\sigma_{\theta}$, and the subscript denotes partial differentiation.  {Note that Lagrange's multiplier is vanished by applying incompressibility condition ($\lambda_{r}=1/(\lambda_{z}\lambda_{\theta}$)}

Following Haughton and Ogden~\cite{haughton1979b}, one solves the equilibrium equation in $r$-direction with the boundary conditions at inner surface ($\sigma_{r}|_{r=r_{i}}=-P$) and outer surface  ($\sigma_{r}|_{r=r_{o}}=0$). Define the internal pressure as
\begin{equation} \label{press}
P=-\int_{\lambda_{i}}^{\lambda_{o}}\frac{\hat{W}_{\theta}}{\lambda_{\theta}^{2}\lambda_{z}-1}d\lambda_{\theta},
\end{equation}
where $\lambda_{i}=\frac{r_i}{R_i}$ and $\lambda_{o}=\frac{r_o}{R_o}$. $\lambda_{i}$ and $\lambda_{o}$ are related by the following equation

\begin{equation}
\lambda_{i}^{2}\lambda_{z}-1=\left(\frac{R_o}{R_i}\right)^{2}(\lambda_{o}^{2}\lambda_{z}-1).
\end{equation}
The resultant end force can be obtained by
\begin{equation} \label{force}
F_e=\pi R_{i}^{2}\left[(1-\lambda_{i}^{2}\lambda_{z})\int_{\lambda_{i}}^{\lambda_{o}}\frac{2\lambda_{z}\hat{W}_{z}-\lambda \hat{W}_{\theta}}{(1-\lambda_{\theta}^{2}\lambda_{z})^2}\lambda_{\theta}d\lambda_{\theta}\right].
\end{equation}
The thin wall assumption (${R_i}>>{H}$) so that $R_i \approx R_o = R$, and the incompressibility condition will lead to $\lambda_{z}rh=RH$ ~\cite{kyr1990}. Consequently, \req{press} and \req{force} will reduce to
\begin{equation} \label{press1}
P=\frac{1}{\lambda_{z}\lambda_{\theta}}\frac{H}{R}\hat{W}_{\theta},
\end{equation}
\begin{equation} \label{force1}
F_e=2\pi RH(\hat{W}_{z}-\frac{\lambda_{\theta}}{2\lambda_{z}}\hat{W}_{\theta}).
\end{equation} The above equations identify the bifurcation of the uniformly inflated tube to a localized bulge state. The bulge profile is to be obtained. A system of ordinary differential equations using thin wall assumption and the finite deformation of the axisymmetric membrane (discussed in~\cite{green1960}) have been derived for  the bulge profile in~\cite{kyr1991} for a prescribed volumetric inflation under quasi-static conditions:  
\begin{equation} \label{asm}
\begin{array}{lc}
\frac{d\lambda_{z}}{dZ}=\frac{1}{R\hat{W}_{zz}}\left[(\lambda_{\theta}\hat{W}_{z\theta}-\hat{W}_{z})\frac{dR}{dZ}+\sin({\omega})(\hat{W}_{\theta}-\lambda_{z}\hat{W}_{z\theta})\frac{dS}{dZ}\right],\\
\frac{d\lambda_{\theta}}{dZ}=\frac{1}{R}\left[\lambda_{z}\sin({\omega})\frac{dS}{dZ}-\lambda_{\theta}\frac{dR}{dZ}\right], \\
\frac{d\omega}{dZ}=\frac{1}{R\hat{W}_{z}}\left[\cos({\omega})(\hat{W}_{\theta}-2\frac{\lambda_{z}}{\lambda_{\theta}}\hat{W}_z)+\frac{F_e}{\pi RH}\frac{\lambda_{z}}{\lambda_{\theta}}\right]\frac{dS}{dZ},\\
\frac{dz}{dZ}=-\lambda_{z}\cos({\omega})\frac{dS}{dZ},	
\end{array}
\end{equation} 
 {where $\omega$ is the angle between the normal of the tube wall and r-axis (see \fref{Fig_1})}, it is assumed that the tube is symmetrical with respect to the plane $Z=0$, before and after the bulge formation. We can use the above set of ordinary differential equations to determine the bulge profile using appropriate boundary conditions. Also,  it is possible to include the effect of geometrical imperfection in the model by introducing an initial nonuniform tube profile. It has been assumed that the strain energy density function (SEF) is defined in terms of principal \emph{thickness averaged} stretches ($\hat{W}(\lambda_{z}, \lambda_{\theta})$). The four boundary conditions
\begin{equation}\label{asmbc}
\omega(0)=0,\quad  z(0)=0, \quad \lambda_{\theta}(L)=1,  \quad \lambda_{\theta}(0)=\displaystyle\frac{D_{max}}{D}, \end{equation}
define the boundary value problem to be solved for the bulge profile, where $\lambda_z$ depends on $\lambda_{\theta}$ via the second differential equation in~\req{asm}.  $Z=0$ at the crown of the bulge and $Z=L$ at the end of the tube. Note that the tube is of an undeformed length $2L$. In the above,  the tube is subjected to a fixed end force (i.e. dead weight). Accordingly, the axial force (end force $F_e$), is a constant parameter in deriving the bifurcation condition associated with uniform expansion to localized bulge. 

In the case of prescribed axial force where dead-weight acts at one end, the bulge can move axially and there is no buckling. There are only four unknowns $(\lambda_z, \lambda_\theta,\omega,z)$ in~\req{asm}, since $F_e$ is prescribed.  Hence, the four boundary conditions in~\req{asmbc}  together with four ODEs in~\req{asm} are sufficient.   However, in our tests under prescribed volumetric inflation and under fixed grips (displacement control), we observe  that the crown of the bulge $Z=0$ does not move axially, but instead the bulge grows in size and then buckles. The above equations in~\req{asm} remain valid; however, the boundary conditions require careful implementation, since  the end force $F_e$ is not constant in~\req{asm} but  a function of $\lambda_{\theta}$ and $\lambda_{z}$ at the stationary end $z(L)=l$. One now has to contend with five unknowns $(\lambda_z, \lambda_{\theta},\omega,z, F_e)$ with an additional condition at $z(L)=l$. The varying $F_e$  is implemented incrementally in the numerical solution using the shooting method.  Starting from the beginning of inflation with an initially known tension $F_e=F_i$, we proceed to the next step  of   incremental expansion  by assigning an incremental  change in the crown diameter $D_{max}$ in $\lambda_{\theta}(0)=\displaystyle {D_{max}}/{D}$ and solve for the remaining four unknowns. Then, using~\req{force1} we find $F_e$ for the next increment. As a side remark, we found that in the MATLAB$^\text{\textregistered}$ ODE toolbox, the  ODE113 solver  was found  to be more accurate compared with ODE45 and ODE23. After solving the ODE set (\ref{asm}) the associated values for internal pressure and end force can be obtained from~\req{press1} and~\req{force1}.

\subsection{Diffuse interface model}
The precise identification of the bifurcation point is sensitive and computationally expensive by solving the four coupled differential equations in~\req{asm} together with the boundary conditions in~\req{asmbc}. Particularly, for the fixed grip case, the end force increment size matters. Recent developments, within the membrane approximation, lead to the postulation of the so called diffuse interface model, in which the four ordinary equations in~\req{asm} are replaced by a single scalar master equation. Besides numerical efficiency, the shooting method can give a more accurate bifurcation point, which motivates  the regularized asymptotic problem in the slender tube limit. 

In the thin wall membrane approximation (${D}>>{H}$), regularized models for slender structures with non-convex $\hat{W}$ can be obtained through an asymptotic expansion by selecting the aspect ratio ${R}/{L}$ as the parameter for expansion~\cite{lest2018,lest2020}.  The three-dimensional axisymmetric membrane model is thus reduced to a one dimensional diffuse interface model by including the circumferential stretch gradient, through the relationship between Green-Lagrange strains and the principal stretches. 
The Green-Lagrange strain tensor is defined in terms of axial $(\lambda_{z})$ and circumferential $(\lambda_{\theta})$  stretches by
\begin{equation}
{\bm{E}}=\left({\begin{array}{c}	E_{\theta} \\ E_{z} \end{array}}\right)=\frac{1}{2}
\left(\begin{array}{c} \lambda_{\theta}^{2}-1 \\ \lambda_{z}^{2}+(R\lambda^{'}_{\theta})^2-1 \end{array}\right),
\end{equation}
where, by separating the gradient term ($\lambda^{'}_{\theta}=\frac{d\lambda_{\theta}}{dZ}$) we have
\begin{equation}
{\bm{E}}={\bm{E}}_{0}(\lambda_{z},\lambda_{\theta})+{\bm{E}}_{1}(\lambda^{'}_{\theta})=\frac{1}{2}
\left(\begin{array}{c} \lambda_{\theta}^{2}-1 \\ \lambda_{z}^{2}-1 \end{array}\right)+\frac{1}{2}\left(\begin{array}{c} 0 \\ (R\lambda^{'}_{\theta})^2 \end{array}\right).
\end{equation}
 {Note that here, the gradient of $R$ are assumed to be small. Hence, $\omega$ in \fref{Fig_1} is close to $0$, $s\simeq z$ and $\lambda_z=ds/dS\simeq dz/dZ$.} The strain energy density function can now be defined in terms of the first term (${\bm{E}}_{0}(\lambda_{z},\lambda_{\theta})$) if one neglects the circumferential stretch gradient ($\lambda_{\theta}^{'}$) as
\begin{equation}
\hat{W}({\bm{E}}_{0})=\hat{W}_{0}(\lambda_{z},\lambda_{\theta})=\sum_{i=1}^{3}\frac{s_i}{\alpha_i}\left( \lambda_{z}^{\alpha_i} + \lambda_{\theta}^{\alpha_i}+ (\frac{1}{\lambda_{z}\lambda_{\theta}})^{\alpha_i }-3 \right),
\end{equation}
where  $\lambda_{z}=\sqrt{E_{z}+1}$ and $\lambda_{\theta}=\sqrt{E_{\theta}+1}$.

 {The total potential energy of the system is}
\begin{equation} \label{totalu}
 {G=\int_{0}^{L}(\hat{W}_{0}2 \pi RH dZ -\pi (R\lambda_{\theta})^{2}\lambda_{z}P dZ-\lambda_{z}F_{e} dZ).}
\end{equation}
Accordingly, the scaled membrane energy per unit reference length is
\begin{equation} \label{memen}
g_0(P^{*},\lambda_{z},\lambda_{\theta})=\frac{1}{S_{ini}}\hat{W}_{0}-P^{*}\frac{e}{2}\lambda_{z}\lambda_{\theta}^{2}-F^{*}\lambda_{z},
\end{equation}
where the initial shear modulus in Ogden's SEF is $S_{ini}=\displaystyle\frac{1}{2}\sum_{i=1}^{3}{s_i}\alpha_i$, the scaled internal pressure $P^{*}={P}/{S_{ini}}$ and the scaled axial force $F^{*}={F_e}/{(2\pi RHS_{ini})}$ and $e={R}/{H}$ is the radius to  the thickness ratio. Note that $F^{*}$ can depend on the boundary conditions and the axial equilibrium through~\req{force1}. The equilibrium conditions follow from  
\begin{equation} \label{lam0}
\frac{\partial g_0}{\partial \lambda_{z}}(P^{*},\lambda_{z},\lambda_{\theta})=0,
\end{equation}
\begin{equation} \label{mu0}
\frac{\partial g_0}{\partial \lambda_{\theta}}(P^{*},\lambda_{z},\lambda_{\theta})=0.
\end{equation}
One can solve equation (\ref{lam0}) to find the equilibrium value for $\lambda_{z}$ which in turn can be expressed as a function of $P^{*}$ and $\lambda_{\theta}$:
\begin{equation} \label{lambdas}
\lambda_z=\lambda_0(P^{*},\lambda_{\theta}).
\end{equation}	
Inserting the above in~\req{memen} and~\req{mu0}, the membrane energy per unit length ($G_0$) and circumferential equilibrium equation  follow as:
\begin{equation}
G_{0}(P^{*},\lambda_{\theta})=g_0(P^{*},\lambda_0,\lambda_{\theta}),
\end{equation}
and
\begin{equation}
n_{0}(P^{*},\lambda_{\theta})=-\frac{\partial g_0}{\partial \lambda_{\theta}}(P^{*},\lambda_0(P^{*},\lambda_{\theta}),\lambda_{\theta})=0.
\end{equation}	
Solving the above reduced equilibrium condition ($n_{0}=0$) gives the homogeneous solution for the expansion of the cylinder due to the internal pressure. To achieve the solution, we start with equation (\ref{lam0}) by tabulating a set of roots in terms of $P^{*}$ and $\lambda_{\theta}$, which gives us $\lambda=\lambda_0(P^{*},\lambda_{\theta})$. Afterwards, by employing the arclength method~\cite{lest2018}, the reduced equilibrium condition is solved using equation (\ref{lambdas}). 

In order to obtain the non-homogeneous solution, we can implement the contribution of the circumferential stretch gradient by assuming ${\bm{E}}_{1}$ ($\mathcal{O}(\epsilon)$) a small correction to ${\bm{E}}_{0}$ ($\mathcal{O}(1)$). The order of magnitudes is inline with the finite elasticity theory and under the assumption that $L>>R$~\cite{lest2018}. The strain energy density function can be expanded as
\begin{equation}
\hat{W}({\bm{E}})=\hat{W}({\bm{E}}_{0}+{\bm{E}}_{1})=\hat{W}_{0}({\bm{E}}_{0})+\frac{\partial \hat{W}}{\partial {\bm{E}}}({\bm{E}}_{0}).{\bm{E}}_{1}+\mathcal{O}(\epsilon^{4}).
\end{equation}
By neglecting the higher order terms ($\mathcal{O}(\epsilon^{4})$), an approximate the total potential energy is
\begin{equation} \label{totalen}
G^{*}=\int_{0}^{L}G_{0}(P^{*},\lambda_{\theta}(Z))dZ+\frac{1}{2}\int_{0}^{L}G_{1}(P^{*},\lambda_{\theta}(Z)){\lambda_{\theta}^{'}}^2(Z)dZ,
\end{equation}
which includes the contribution of the membrane energy per unit length $G_0$ and the contribution due to stretch gradient  $ G_{1}$. The stretch gradient contribution evaluates to
\begin{equation}
G_{1}(P^{*},\lambda_{\theta}(Z))=R^2\left[\frac{1}{\lambda_0}\frac{1}{S_{ini}}\frac{\partial \hat{W}_{0}}{\partial \lambda_z}(\lambda_z,\lambda_{\theta}(Z))\right]_{\lambda_z=\lambda_0(P^{*},\lambda_{\theta}(Z))}.
\end{equation}
It should be noted that in deriving equation (\ref{totalen}), we replace the axial stretch ($\lambda_{z}$) with the axial stretch function (equation (\ref{lambdas})) from the homogeneous solution. Accordingly $\lambda_{z}(Z)$ can be approximated using the scaled internal pressure ($P^{*}$) and the local circumferential stretch ($\lambda_{\theta}(Z)$).
The non-linear equilibrium condition is obtained from equation (\ref{totalen}), with $\lambda_{\theta}^{'}(z=0)=\lambda_{\theta}^{'}(z=l)=0$ as boundary conditions, using the Euler-Lagrange method
\begin{equation} \label{dimeq}
n_{0}(P^{*},\lambda_{\theta}(Z))-\frac{1}{2}\frac{\partial G_{1}}{\partial \lambda_{\theta}}(P^{*},\lambda_{\theta}(Z)){\lambda_{\theta}^{'}}^{2}(Z)+\frac{d}{dZ}(G_{1}(P^{*},\lambda_{\theta}(Z))\lambda_{\theta}^{'}(Z))=0.
\end{equation}

The above equation can be solved using quadrature or direct numerical  integration method to find the relationship between $\lambda_{\theta}$ and $P^*$.  Assuming $\lambda_{\theta}(Z)={\lambda}_{\theta}+\epsilon(Z)$ a perturbation from homogeneous solution $\lambda_{\theta}$, the linear bifurcation analysis, with $\lambda_{\theta}^{'}(z=0)=\lambda_{\theta}^{'}(z=l)=0$ as boundary conditions, yields the bifurcation condition
\begin{equation} \label{bcdim}
\frac{\partial n_{0}}{\partial \lambda_{\theta}}(P^{*},\lambda_{\theta})=G_{1}(P^{*},\lambda_{\theta})\frac{\pi^2}{L^2},
\end{equation}
which has to be solved along with the circumferential equilibrium~\req{mu0}. 

 {For a fixed-fixed boundary condition, the end force does not contribute to the total potential energy during inflation. Consequently, $G$ in \req{totalu} reduces to

\begin{equation} \label{newtotalu}
G=\int_{0}^{L}(\hat{W}_{0}2 \pi RH dZ -\pi (R\lambda_{\theta})^{2}\lambda_{z}P dZ).
\end{equation}

The effect of boundary conditions can be introduced by defining the constraint on the length of the tube as

\begin{equation} \label{lencon}
\int_{0}^{L}(\lambda_{z} dZ)-l=\int_{0}^{L}(\lambda_{z}-\frac{l}{L})dZ=0.
\end{equation}

Not that this is the integral form of differential equation $dz/dZ=\lambda_{z}$ with $z(Z=o)=0$ and $z(Z=L)=l$ as boundary conditions. The Lagrangian of the above system of \req{lencon} and \req{newtotalu} with $F^\dagger$ reads

\begin{equation}
	\mathcal{L}=\int_{0}^{L}(\hat{W}_{0}2 \pi RH dZ -\pi (R\lambda_{\theta})^{2}\lambda_{z}P dZ-F^{\dagger}[(\lambda_{z}-\frac{l}{L})dZ]).
\end{equation}

In the view of the Lagrangian, the new form of scaled membrane energy per unit length is in the similar form of \req{memen}

\begin{equation}
g_0^{\dagger}(P^{*},\lambda_{z},\lambda_{\theta})=\frac{1}{S_{ini}}\hat{W}_{0}-P^{*}\frac{e}{2}\lambda_{z}\lambda_{\theta}^{2}-F^{*}(\lambda_{z}-\frac{l}{L}),
\end{equation}
Where $F^{*}=F^{\dagger}/{(2\pi RHS_{ini})}$ is the scaled Lagrange's multiplier. 
Differentiation with respect to $\lambda_{z}$ and $\lambda_{\theta}$ reads the same for of axial \req{lam0} and circumferential \req{mu0} equilibria. Differentiation with respect to $F^{\dagger}$, the Lagrange's multiplier, reads $\lambda_{z}-l/L=0$, which leads to trivial solution $\lambda_{z}(Z)=l/L$. To avoid the trivial solution in numerical analysis, we use the alternative differential equation
\begin{equation} \label{const}
dz/dZ=\lambda_{Z},
\end{equation}

where, $z(Z=o)=0$ and $z(Z=L)=l$ impose the boundary conditions at fixed ends. The axial equilibrium \req{lam0} should be solved with \req{const} to get the new form of \req{lambdas} associated with fixed-fixed boundary conditions. The diffuse interface model \req{totalen} and bifurcation condition \req{bcdim} will have the same form as before. However, they have to be solved along with the additional constraint condition \req{const}.

}

\subsection{Buckling analysis}
\begin{figure}[!h]
	\center
	\includegraphics[width=0.6\textwidth]{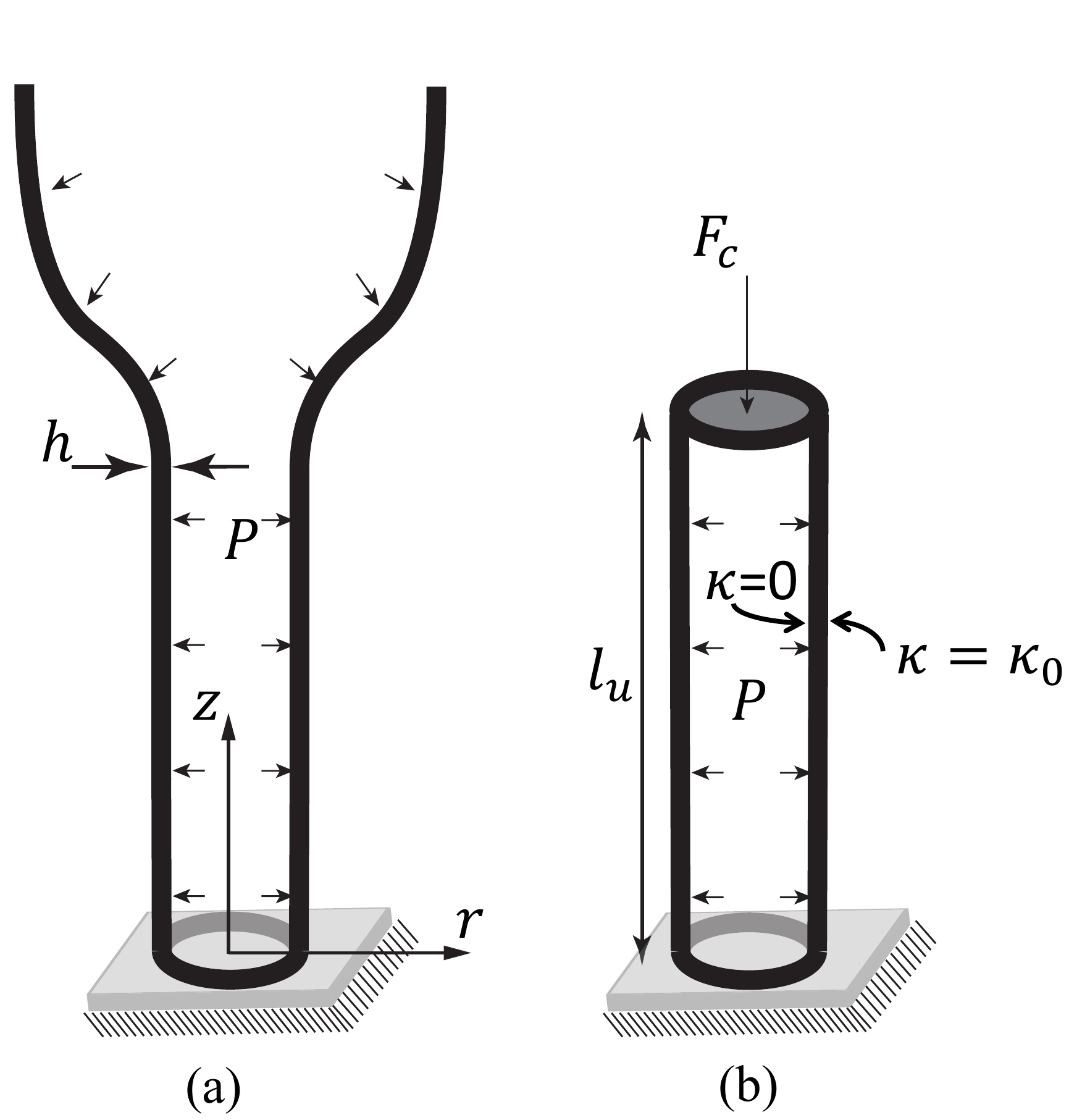}
	\caption{ (a) Schematic  for the buckling analysis of the uniform section of the tube of length $l_{u}$. (b) the decoupled uniform section is considered as a tube subjected to axial compressive force $F_c$ and internal pressure $P$.}
	\label{Fig:5}
\end{figure}

In this section, we study the lateral buckling of a tube subjected to an axial force and an internal pressure under fixed-fixed boundary conditions using an incremental hyperelastic formulation. In order to simplify the analysis, we assume a uniform tube without the bulge, by decoupling buckling and bulge phenomena. This approximation is not valid in general but applicable in our case. In practice, buckling follows bulge formation. However,  it is observed in  our experiments that lateral buckling deformation occurs in the uniform region of the tube, presumably due to the increased bending resistance of the bulged section.  The conditions shown in~\fref{Fig:5} are assumed to hold.  Note that the analysis to follow does not account for twist-flexure coupling, which can be important if the interest is in post buckling response. The present analysis is thus limited to the onset of buckling, and is not capable of predicting the post buckling shapes. Also, here we drop the membrane assumption in deriving the buckling condition. In the numerical evaluation,  the thickness averaged  stretches with the thickness averaged stretches ($\lambda_{z}$ and $\lambda_{\theta}$) can be used. The incompressibility condition is through a Lagrange multiplier. To conduct the buckling analysis, we follow~ \cite{goriely2008} and~\cite{Chen2017}. We assume ${\bm{W}}=W{({\bm{F}})}$ as the strain energy density function for a hyperelastic material, where ${\bm{F}}$ is the deformation gradient. The stress-deformation relation is then given by
\begin{equation} \label{19}
{\bm{S}}=\frac{\partial {\bm{W}}}{\partial {\bm{F}}}-\Gamma{\bm{F}}^{-1}.
\end{equation}
\begin{equation}
{\bm{\sigma}}={\bm{F}}\frac{\partial{\bm{W}}}{\partial {\bm{F}}}-\Gamma\bm{I}.
\end{equation}
where $\Gamma$ is the Lagrange multiplier associated with the incompressibility of hyperelastic materials, $\bm{S}$ is nominal stress tensor, and $\bm{\sigma}$ is the Cauchy stress tensor. Equilibrium equations for the incremental deformation field are~\cite{ogden1997}
\begin{equation} \label{21}
div\dot{{\bm{S}}}_{0}=0,
\end{equation}
where $\bm{\dot{S}_0}$ is the incremental nominal stress tensor. As the first-order approximation of the constitutive relations, $\bm{\dot{S}_ 0}$ is 
\begin{equation}
\dot{{\bm{S}}}_{0}={\bm{B}}{\bm{\eta}}+\Gamma{\bm{\eta}}-\dot{\Gamma}\bm{I},
\end{equation}
where $\dot{\Gamma}$ is the increment in Lagrange multiplier,  $\bm{\eta}=grad(u)$, $u$ is the displacement field, and  ${\bm{B}}$ is the fourth-order tensor of instantaneous elastic moduli. $\bm{I}$ is the unit tensor. The components of $\bm{B}$ are given by:
\begin{equation} \label{23}
{{B}}_{piqj}={{F}}_{p\alpha}{{F}}_{q\beta}\frac{\partial^{2} {W}_{i\alpha j\beta}}{\partial {{F}}_{i\alpha}\partial {{F}}_{j\beta}} ~.
\end{equation}
For the deformation  increment we have
\begin{equation} \label{24}
{\bm{\eta}}=\left(
\begin{array}{ccc}
\frac{\partial u_r}{\partial r} & \frac{1}{r}(\frac{\partial u_r}{\partial \theta}-u_\theta) & \frac{\partial u_r}{\partial z} \\
\frac{\partial u_\theta}{\partial r} & \frac{1}{r}(\frac{\partial u_\theta}{\partial \theta}+u_r) & \frac{\partial u_\theta}{\partial z} \\
\frac{\partial u_z}{\partial r} & \frac{1}{r}\frac{\partial u_z}{\partial \theta} & \frac{\partial u_z}{\partial z}
\end{array}\right).
\end{equation} 
Also, the incremental incompressibility condition is
\begin{equation} \label{25}
\frac{\partial u_r}{\partial r}+\frac{1}{r}(\frac{\partial u_\theta}{\partial \theta}+u_r)+\frac{\partial u_z}{\partial z}=0.
\end{equation}
Based on the equilibrium at inner surface in the normal direction of the traction vector, the boundary conditions at the inner surface are
\begin{equation} \label{26}
\dot{{{S}}}_{0rr}=P\frac{\partial u_r}{\partial r},~\dot{{{S}}}_{0r\theta}=P\frac{1}{r}(\frac{\partial u_r}{\partial \theta}-u_\theta),~\dot{{{S}}}_{0rz}=P\frac{\partial u_r}{\partial z}.
\end{equation}

Defining $\Sigma_{ij}=r\dot{S}_{0ij},\nabla_{r}=r{\partial}/{\partial r}$, we can form a state space system~\cite{chen2005} using equations~\req{19},~\req{21},~\req{24}, and~\req{25}:
\begin{equation} \label{27}
\nabla_{r}{\bm{U}}={{\bm{M}}{\bm{U}}},
\end{equation}
where ${\bm{U}}=[u_{r}, u_{\theta}, u_{z}, \Sigma_{rr}, \Sigma_{r\theta}, \Sigma_{rz}]^{T}$ is the state vector, with $u_{r}, u_{\theta}, u_{z}, \Sigma_{rr}, \Sigma_{r\theta}, $ and $ \Sigma_{rz}$ being the state variables, and  $\bm{M}$ is the system matrix of order $6\times6$, which contains partial derivatives with respect to $\theta$ and $z$ only.  Note  here that the origin $Z=0$ is not at the crown of the bulge but is located at one of the ends $L=l_{u}$ as shown in~\fref{Fig:5}).	
Assuming a smooth contact at  both ends of the tube $(z=0,L)$, the boundary conditions on the incremental traction become
\begin{equation} \label{28}
{T}_{r}(z)=0,~{T}_{\theta}(z)=0,~u_z(z)=0, \text{ at~} z=0, \quad z=L,
\end{equation}
where $T_{r}(z)$ and $T_{\theta}(z)$ are the incremental traction vector components in the polar co-ordinates, and $u_z(z)$ is the incremental axial displacement. 
To satisfy the above six boundary conditions in~\req{28} and~\req{26}, we assume the following displacement vector in the cylindrical polar co-ordinates ($r$,$\theta$,$z$) shown in \fref{Fig:5} 
\begin{equation}\label{assumedU}
{\bm{U}}=\left(\begin{array}{c}
au_{r}(\rho)\cos(m\theta)\cos(n\pi\zeta) \\
au_{\theta}(\rho)\sin(m\theta)\cos(n\pi\zeta) \\
au_{z}(\rho)\cos(m\theta)\sin(n\pi\zeta) \\
p_{0}a\Sigma_{rr}(\rho)\cos(m\theta)\cos(n\pi\zeta) \\
p_{0}a\Sigma_{r \theta}(\rho)\sin(m\theta)\cos(n\pi\zeta) \\
p_{0}a\Sigma_{rz}(\rho)\cos(m\theta)\sin(n\pi\zeta) \\
\end{array}\right),
\end{equation}
where $a$ is the amplitude; $m=0, 1, 2, ...$ denote the circumferential mode number; and $n=0,1,  2, ...$  is the axial mode number;
$p_{0}$ is a pressure like quantity, having the unit of elastic modulus (i.e. $N/m^2$); $\rho={r}/{r_{i}}$ and $\zeta={z}/{L}$ are the dimensionless radial and axial variables, respectively. The equilibrium equation (\ref{27}) then turns into
\begin{equation} \label{30}
\nabla_{\rho}{\bm{U}}(\rho)={{\bm{M}}}(\rho){\bm{U}}(\rho),
\end{equation}
where $\nabla_{\rho}=\rho{\partial}/{\partial\rho}$. To reduce the computation time and improve convergence, we can adopt $\rho=\exp(\kappa)$ following~\cite{Chen2017}. Accordingly,~\req{30} becomes
\begin{equation} \label{31}
\begin{array}{cc}
\frac{\partial}{\partial\kappa}{\bm{U}}(\kappa)={\bm{M}}(\kappa){\bm{U}}(\kappa) & (\kappa\in[0,\kappa_0]), \kappa_0=ln(\frac{r_{o}}{r_{i}}).
\end{array}
\end{equation}
The elements of ${\bm{M}}$ are given in Table~\ref{elm} in  Appendix B. By discretizing the tube wall  uniformly into $N$ number of layers, each of thickness $t$, we can write $\kappa_{0}=Nt$. Hence, for the $j$th layer ($j\in[1,N]$) $\kappa_{j}=jt$. By employing a layer-wise method for laminated inhomogeneous tubes we can relate state variables at the inner wall to the corresponding ones at the outer wall~\cite{fan1992} as
\begin{equation}\label{32}
{\bm{U}}(\kappa_0)={\bm{D}}{\bm{U}}(\kappa_j=0),
\end{equation}
where ${\bm{D}}=\Pi^{1}_{j=N}exp({\bm{M}(\kappa_{j})}t)$ and $\Pi$ is the product operator. Note that $\kappa_j=0$ corresponds to the inner surface $r=r_i$, and $\kappa_0=Nt$ corresponds to the outer surface $r=r_o$. In the absence of external pressure, the boundary conditions at the inner and the outer surface read
	\begin{equation} \label{33}
	\begin{array}{c}
	\Sigma_{rr}=\frac{-P}{p_{0}}(u_{\theta}+u_{r}+\alpha u_{z}), \Sigma_{r\theta}=\frac{-P}{p_{0}}(mu_{r}+u_{\theta}), \Sigma_{rz}=\frac{-P}{p_{0}}\alpha u_{r} \quad (\kappa_j=0,\text{~inner~surface}), \\
	\Sigma_{rr}=0,~\Sigma_{r\theta}=0,~\Sigma_{rz}=0 \quad (\kappa_j=\kappa_{0}=Nt,\text{~outer~surface}),
	\end{array}
	\end{equation}
	where $\alpha={n\pi a}/{L}$. The above boundary conditions can also be expressed in terms of the displacement by using~\req{33} and the resulting set of linear algebraic equations can be assembled into the form
	\begin{equation} \label{lbc}
	{\bm{C}}~[u_{r}(\kappa_j=0), u_{\theta}(\kappa_j=0), u_{z}(\kappa_j=0)]^{T}=\bm{{0}} \Rightarrow \det(\bm{C})=0.
	\end{equation} 
	Elements of the matrix ${\bm{C}}$ are expressed in terms of $m,~n,~P,~\lambda_{z},~\lambda_{\theta}$ (see table \ref{celm} in Appendix B). Here, we consider $m=0$ and $n=1,~2,...,$ to obtain the bifurcation condition for the lateral mode observed in the experiment. Solving the bifurcation condition (equation (\ref{lbc})) for the applied internal pressures one obtains a pressure contour plot in $\lambda_{\theta}$-$\lambda_{z}$ plane identifying  the critical pressures that cause buckling for any chosen point in the $\lambda_{\theta}$-$\lambda_{z}$ plane. 
	
	The analysis can be summarized as follows in two steps: (1) obtain the  bulge profile of $\lambda_{\theta}$ and ${\lambda_{z}}$ is obtained as a function of maximum bulge diameter ($D_{max}/D$) and internal pressure ($P$) using~\req{asm} and~\req{dimeq}; (2) use equation (\ref{lbc}) to check the buckling condition for the internal pressure  and the  $\lambda_{z}$ and $\lambda_{\theta}$ in the uniform region of  the bulged tube. {Note that $l_{u}$ is updated in each step to account for the change in the length of the uniform section of the tube due to bulge propagation}. The assumed uniform circumferential and axial stretches in~\req{lbc} hold only in the uniform region, in~\fref{Fig:5}.

\section{Results and Discussion}
\begin{figure}[!h]
	\center
	\includegraphics[width=0.8\textwidth]{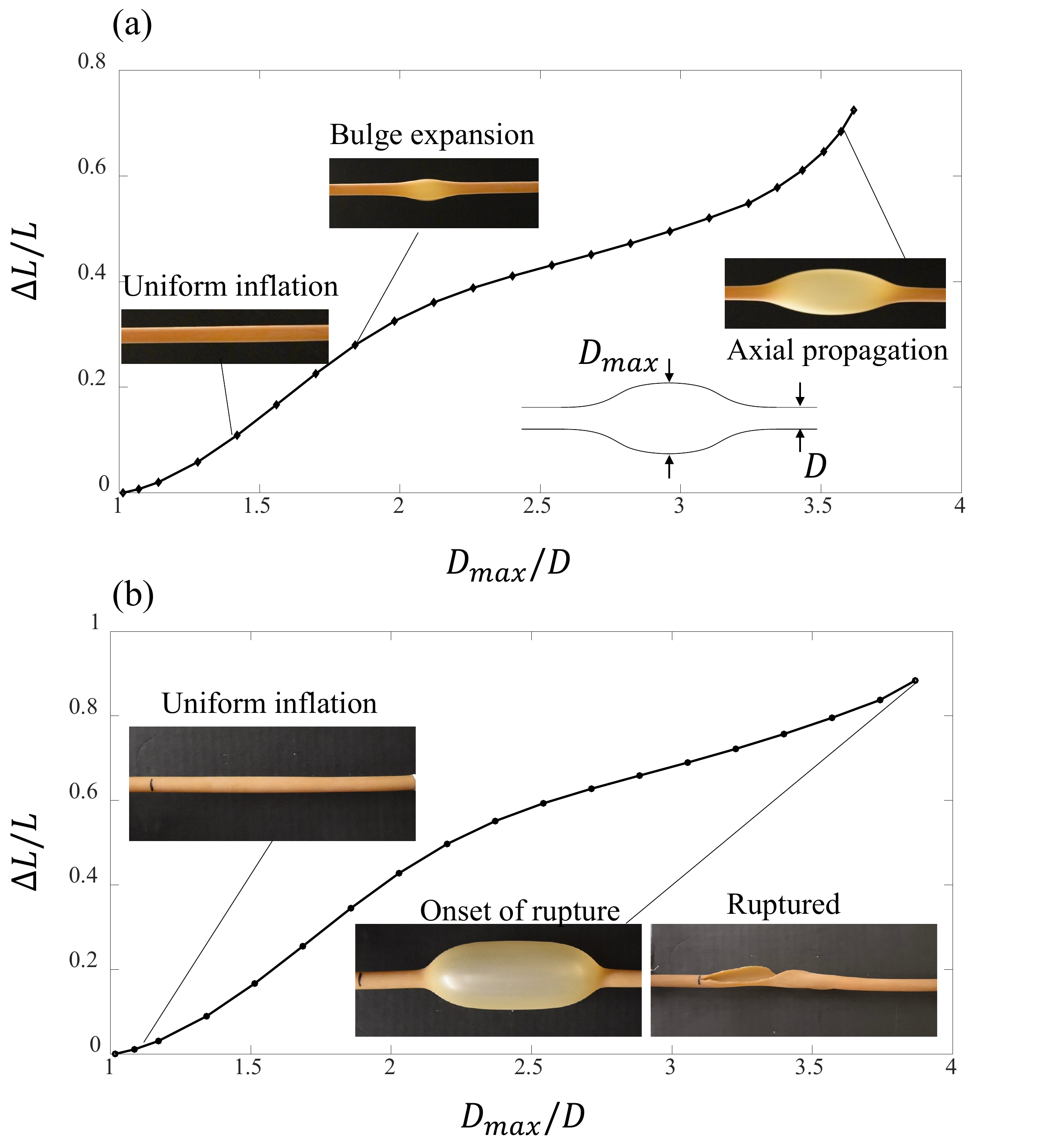}
	\caption{ Deformation stages in a force control test (Fig.\ref{Fig:2}): (i) Uniform inflation, (ii) Bulge expansion (radial propagation), (iii) Axial propagation or rupture. The added mass is $M=800~g$  in (a) and the bulge propagates axially without rupture. In (b) the added mass is $M=3500~g$ and the rupture occurs during axial propagation. (Color online)}
	\label{Fig:4}
\end{figure}

This section compares and contrasts the results from experiments and the analyses. Consider the fixed-free test shown in~\fref{Fig:2} where the tube is fixed at one end and a mass $M$ is attached at the other end to impart initial tension. Quasi-static inflation of the tube shown  in~\fref{Fig:4} begins with a uniform expansion stage, followed by the initiation of the bulge. The bulge  first expands mostly radially growing in diameter and then axially growing in length.  The engineering axial strain is plotted against the circumferential strain measured at the maximum diameter section in \fref{Fig:4}(a). The insets show different stages of inflation, mentioned earlier.  It should be noted that the increase in length with progressive inflation is significant once the bulge is initiated around {${D_{max}}/D=1.4$}. During the propagation stage, there is a considerably high increase in length for almost constant circumferential strain. For the test shown, the end stage is axial yielding without rupture, since the applied initial tension is small. If one increases the tension, the qualitative picture remains the same except that at high circumferential strains the rupture occurs.  For higher values of initial weight (or pre-stretch) the bulge leads to rupture as shown in {~\fref{Fig:4}(b)}.

\begin{figure}[!h]
	\center
	\includegraphics[width=0.7\textwidth]{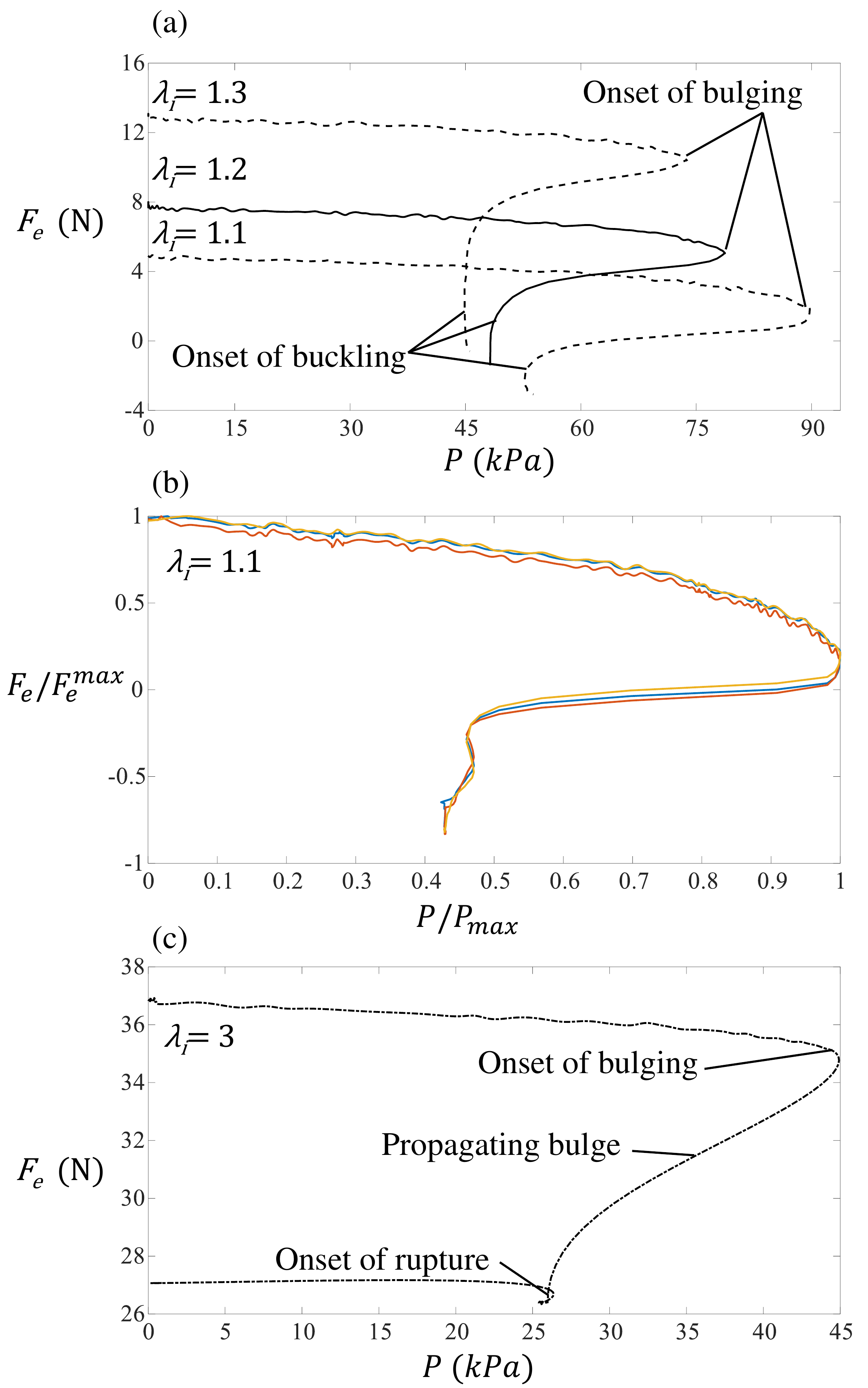}
	\caption{Bulging and buckling under displacement control: data from experiments ($P$ and $F_e$ are defined in Fig.\ref{Fig:1}): (a) Different values of pre-stretch ($\lambda_{i}$), (b) Three different samples with an identical pre-stretch, (c) Force-pressure curve associated with the rupture at high pre-stretch. (Color online)}
	\label{Fig:3}
\end{figure}

The inflation test data is shown in~\fref{Fig:3} for a displacement control test (see~\fref{Fig:1}). The  onset of bulging at a peak pressure is followed by rapid unloading first and then buckling in these tests, as identified in the force-pressure plots. Increasing the initial axial pre-stretch $\lambda_i$ in the tube in~\fref{Fig:3}(a) leads to a significant decrease in the critical pressure for bulging and a somewhat reduced propagation pressure. Repeatability of experiments across three different specimens can be ascertained from~\fref{Fig:3}(b). In these series of tests where the initial pre-stretch is low, the buckling protects against rupture as a fail safe mode.  It is also possible to observe bulging and rupture at higher values of pre-stretch $\lambda_i$ as shown in~\fref{Fig:3}(c).  

\begin{figure}[!h]
	\center
	\includegraphics[width=0.8\textwidth]{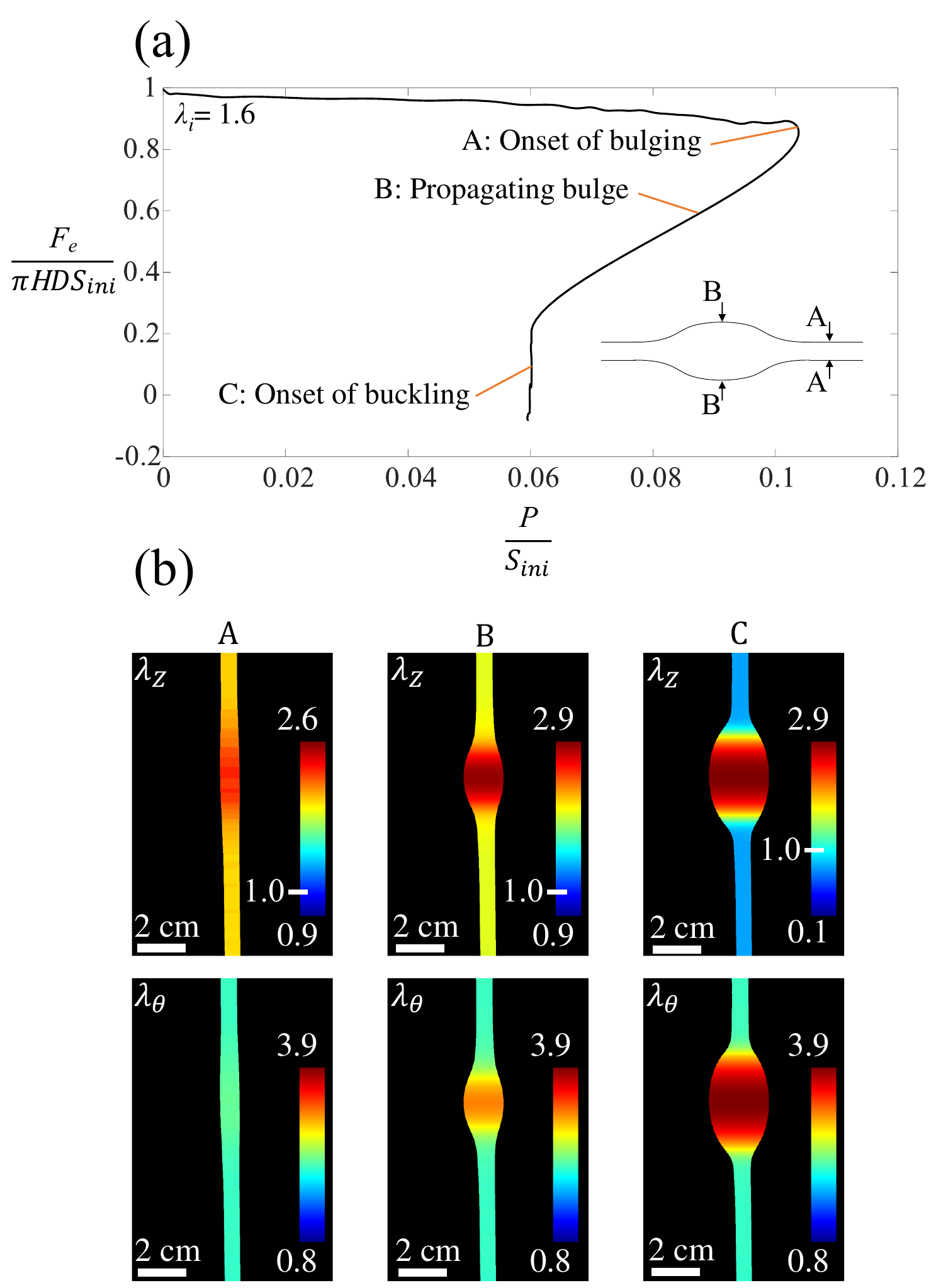}
	\caption{(a) Axial tension-pressure data for displacement control test with bulging and buckling with three stages identified as A, B and C for digital image correlation (DIC) analysis. (b) DIC contours for axial stretch ($\lambda_z$) in the top row and the circumferential stretch ($\lambda_\theta$) in the bottom row. The labels A, B,C correspond to the points identified in (a). Notice that compressive stress ($\lambda_z<1$) emerges at C in the top row outside the bulged section. (Color online)}
	\label{Fig:6}
\end{figure}

\fref{Fig:6} shows the pressure variation during the inflation of the tube subjected to fixed-fixed boundary conditions. As discussed above, the bulge forms when the pressure reaches its maximum value. The entire deformation process is video recorded and the frames of this video are used for subsequent image analysis. For the three different stages indicated in (\fref{Fig:6}a), the axial and circumferential stretches have been calculated by performing digital image correlation (DIC)~\cite{blaber2015ncorr} in MATLAB (see \fref{Fig:6}b). At the onset of bulge formation (point A) in~\fref{Fig:6}(b), we see the localization of the axial stretch at the middle section of the pre-stretched tube, while the entire length of the tube is still under tension ($\lambda_{z}>1$). Point B shows the early stages of the propagation where the axial stretch localization is more intense at the bulged section and leads to relaxation in the uniform section (A-A) shown in~\fref{Fig:6}(a). As the bulge propagates axially (point C), the bulged section (B-B) sees a higher tensile stress while the uniform section undergoes compressive stress ($\lambda_{z}<1$). The build-up of the compressive stress in the uniform section can lead to buckling if the initial pre-imposed axial stretch is overcome and the effective length under compression is large. Thus, buckling will not occur if there is sufficient initial tension, or if the uniform region under compression is short. In the biomedical problem of aneurysms, the axial tension of arteries changes with age and large arterial wall tension can lead to aneurysm rupture without buckling. Similarly, lower wall tension can produce buckling, reducing the rupture risk significantly.

\begin{figure}[!h]
	\center
	\includegraphics[width=0.7\textwidth]{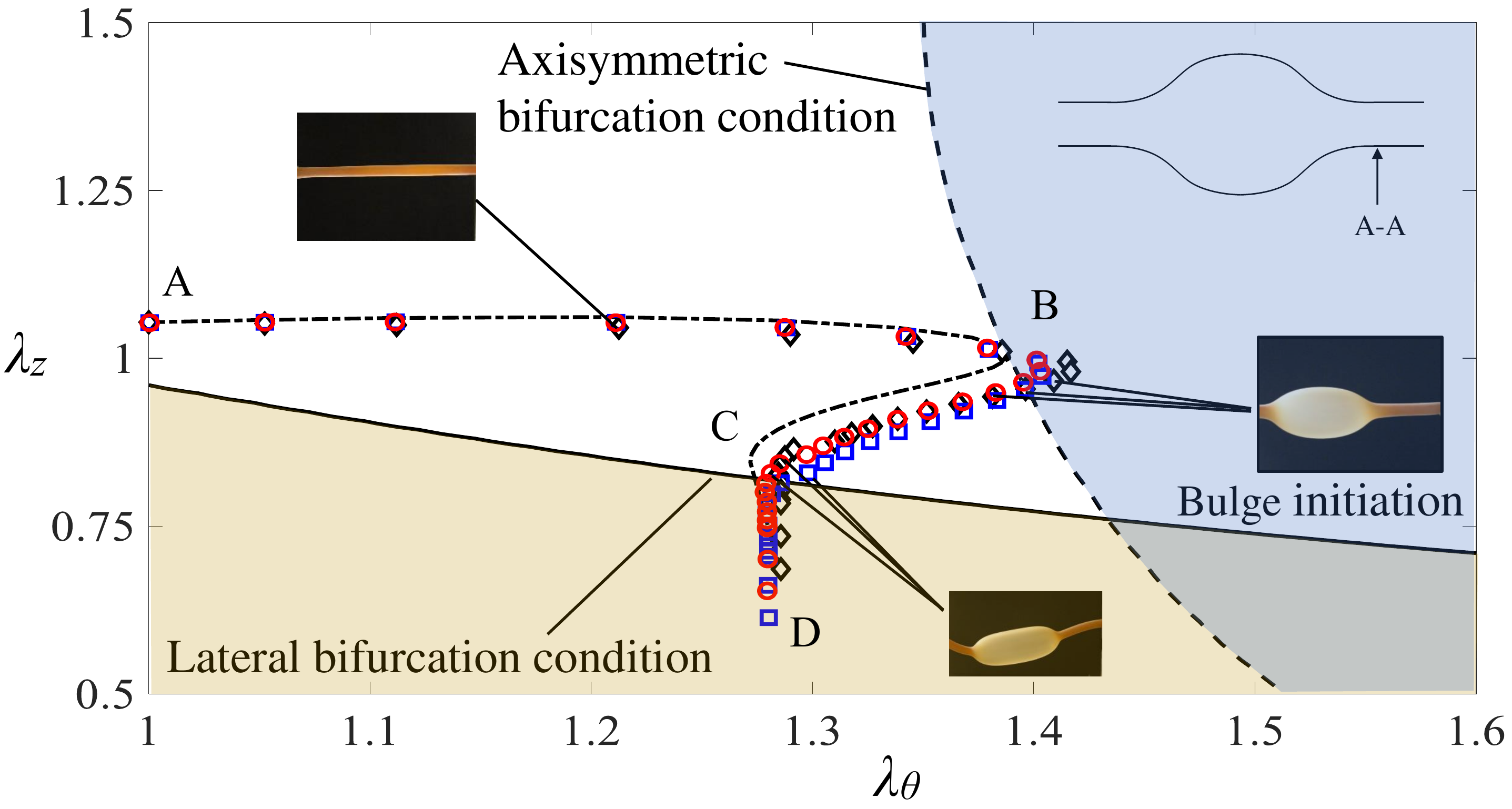}
	\caption{Failure map for displacement control: Dotted line shows the behaviour of the intact section (A-A) during inflation and bulging on the plane of principal stretches. The experimental data points are shown along the analytical dotted line. The blue region (separated by dashed line) shows the bulging regime and the yellow region (separated by solid line) is the buckling regime. When dotted line reaches the bulging regime, it experiences compressive force induced by bulged section. The compressive force directs A-A to the buckling regime. (Color online)}
	\label{Fig:7}
\end{figure}

It is  possible now to compare experiments with the bulging and buckling analyses from Section~\ref{s:analysis}. The regions demarcating the buckling and bulging are identified as shaded regions in the principal-stretch ($\lambda_\theta$-$\lambda_z$) plane, measured at the uniform section A-A, are shown in~\fref{Fig:7}. The analytical trajectory (dash-dotted line)  for the bulge initiation has been constructed based on the axisymmetric membrane analysis using~\req{asm} and the bulge initiation is based on DIM using~\req{bcdim}.   Repeating this analysis for multiple values of the pre-stretch will give the locus for the critical values of  principal stretches associated with bulging. The resulting curve is shown as the dashed line which demarcates the bulging region. Similarly, the  bifurcation condition (solid line) for lateral buckling is constructed based on~\req{lbc}. The parameters listed in Table.~1 are used to construct these analytical curves. Experimental data for the principal stretches for three different specimens are also shown. DIC analysis is used to determine $\lambda_z$ and $\lambda_\theta$, all with the same initial pre-stretch $\lambda_{i}=1.05$. From A to B, we have the uniform inflation and the experimental data points agree with the analytical prediction. Point B, is the onset of bulging which occurs when the trajectory hits the bulging  condition (dashed line). Note that the experimental data points show a slightly larger $\lambda_{\theta}$ at the bifurcation point which may be due to the fact that analytical prediction is stiffer as membrane approximation is used. After the bulge forms, the uniform section (A-A) relaxes and as we saw in \fref{Fig:6}, the compressive stress builds up and $\lambda_z$ decreases. The buckling occurs at a lower compressive axial stretch $\lambda_{z}$, presumably due to imperfect boundary conditions.

\begin{figure}[!h]
	\center
	\includegraphics[width=0.7\textwidth]{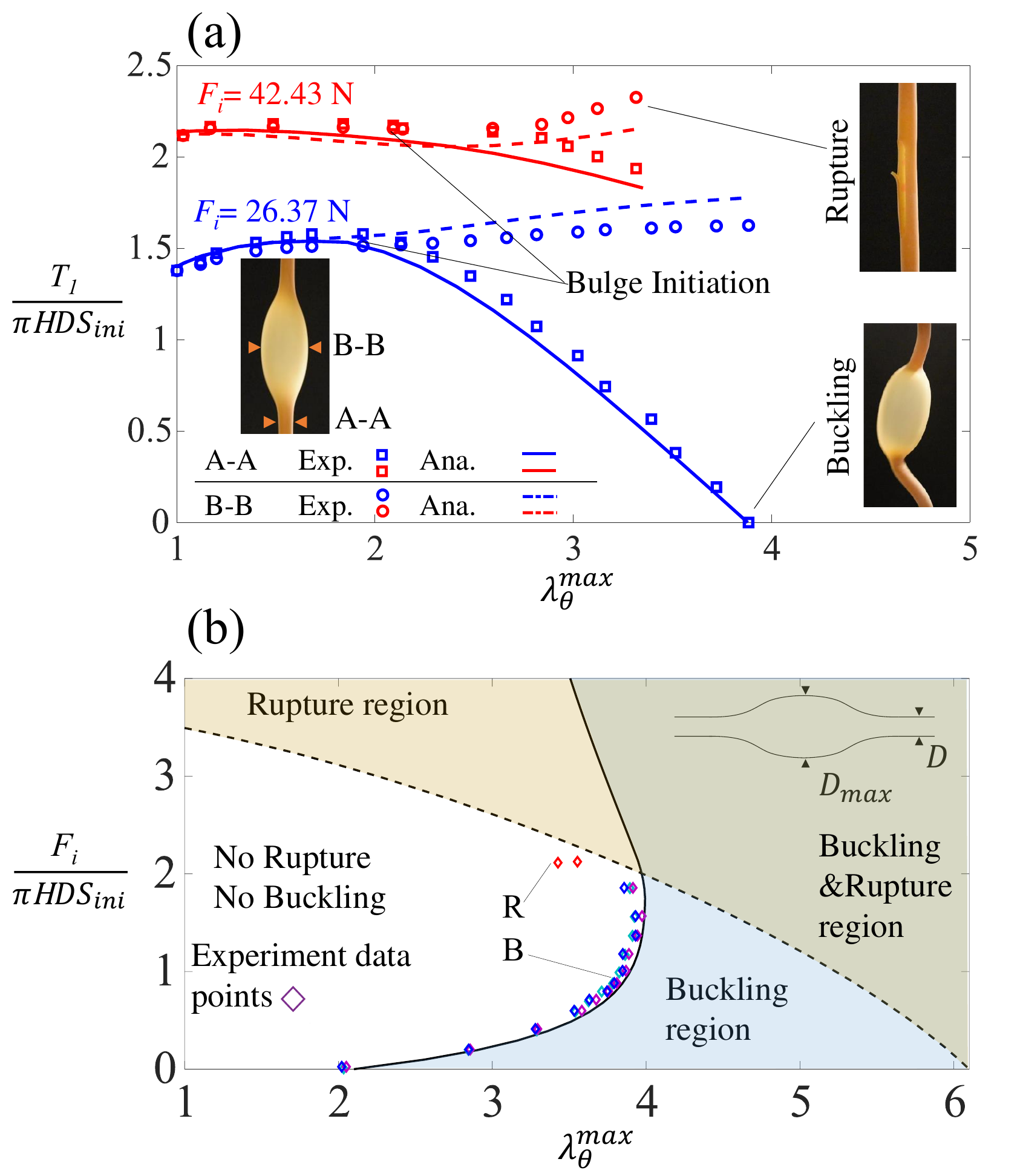}\\
	\caption{{(a) Axial membrane tension ($T_{1}=2\pi r h\sigma_1$)  changes at sections A-A (open square markers) and B-B (open circular markers) until the bifurcation occurs due to bulge formation. Accordingly, the membrane axial tension increases in section B-B and decreases in A-A. In case of rupture (point R in (b)) the tension is preserved in A-A, however, in buckling case (point B in (b)) A-A loses the tension and buckles due to  compressive stresses. (b) Rupture-Buckling failure map. The experimental data points show $\lambda_{\theta}^{max}={D_{max}}/{D}$ at the onset of rupture or buckling. Rupture and buckling zones are identified by yellow and blue regions, respectively. The failure map in (b) shows that: (i) for a buckled bulge, the rupture is observed at a higher diameter ratio, (ii) by increasing the initial axial tension,  rupture occurs without buckling}. (Color online)}
	\label{Fig:8}
\end{figure}

The uniform section (A-A) and bulged section (B-B) have dissimilar behavior in the post bulge formation regime as shown in~\fref{Fig:8} for two different pre-tensions. The membrane axial tension ($T_1=2\pi r h\sigma_1$) has been normalized using the initial shear modulus ($S_{ini}$), tube length, and diameter.  The differences can be observed in the deviation of membrane axial tension during the bulge expansion at these two sections as shown in~\fref{Fig:8}(a). At section A-A outside the bulge, the membrane tension is relieved by bulging  and if the tensions $F_i$ is small, buckling sets in and relieves the axial tension. However, if the initial tension is large, for $F_i=42.34$~N, the initial decrease  in the membrane tension due to bulging is not significant, and this lack of stress relief leads to rupture as there is no buckling. Note that at Section B-B in the bulged region the membrane tension increases, as it should due to the inflation, and hence consistent with the growing radial size of the bulge. The initial tension, $F_i$, is thus a critical parameter that governs the fate of the bulge.  This observation leads to the construction of the failure map shown in~\fref{Fig:8}(b) which portrays the influence of $F_i$. Four distinct regions exist for the post bulge response and they are as identified. At higher values of $F_i$ rupture is the fate of a bulge as the tube is inflated, or $\lambda^{max}_\theta$ is increased.   The rupture criterion is based on the SEF value at ultimate stretch ($\lambda_{ut}$) in the uniaxial tensile test ~\cite{hamdi2006,gent2005}. These quantities are measured experimentally. According to \cite{hamdi2006}, failure energy during the biaxial loading is half of the maximum energy density of the uniaxial tensile test $\hat{W}_{max}=7.50 {J}/{mm^3}=0.5\hat{W}(\lambda_{ut})$.  Using this, we plot the rupture envelope in $F_i$-$\lambda^{max}_{\theta}$, shown as  the dashed line in~\fref{Fig:8}(b).  The solid line  shows the lateral buckling envelope based on~\req{lbc}. The points shown for the experiments correspond to the end points of the test. The test is terminated when buckling or rupture occurs. Note that each experimental data point corresponds to different initial tensions $F_i$, and there are 38 experiments with two rupture cases.  It can be concluded that the calculated envelopes for buckling and rupture are in satisfactory agreement with experiments. The  post buckling can involve twisting of the buckled and bulged tube. This twisting is evident in~\fref{Fig:8} and is beyond the scope of the present analysis. 

\begin{figure}[!h]
	\center
	\includegraphics[width=0.5\textwidth]{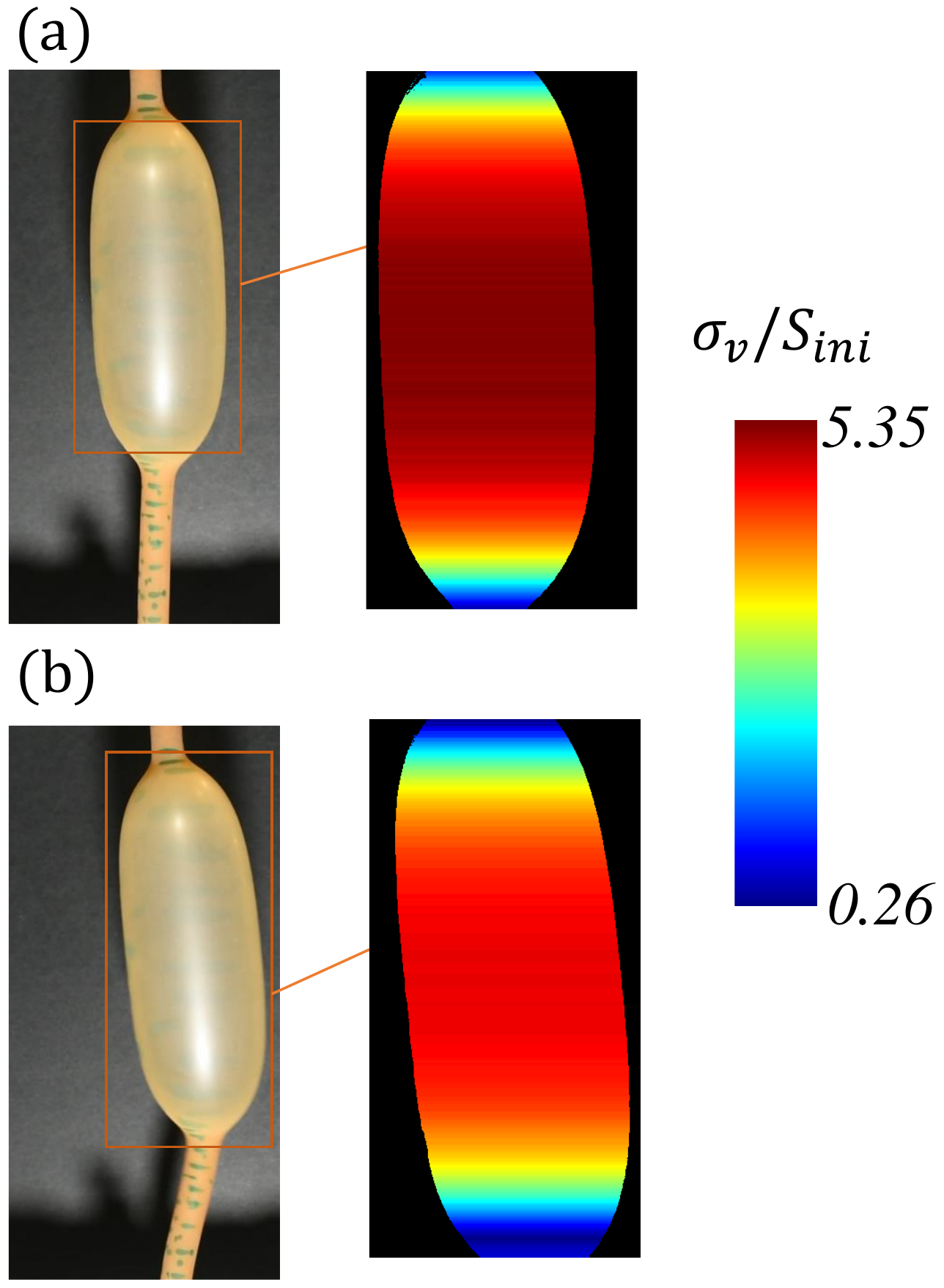}
	\caption{von Mises stress for (a) onset of buckling and (b) early stage of buckling}
	\label{Fig:9}
\end{figure}

 {We can show the stress relief due to buckling by comparing the von Mises stress on the bulged section at the onset of buckling and the early stage of buckled state (see \fref{Fig:9}). The principal stretches are calculated through the DIC method explained above. The stretches are calculated in the plane of buckling and the stretch distribution due to buckling has been approximated using the shape modes in \req{assumedU}. Using the strain energy function, we obtain the principal stresses and von Mises stress. As shown in the \fref{Fig:9}, the peak stress at the crown of the bulge reduces once the buckling occurs.} 

\section{Concluding Remarks}
The fate of a bulge initiated in an inflated cylindrical hyperleastic (latex rubber) tube has been  examined through experiments and analyses. Within the membrane approximation  limit, a justified assumptions for the experiments reported here, the bulge can either buckle  or rupture, depending on (a) boundary conditions and (b) pre-existing axial stretch or tension.  The diffuse interface model and axisymmetric membrane analyses, coupled with lateral buckling analysis produces a portrait spanned by  the axial tension and circumferential strain axes demarcating the dangerous rupture and fail safe buckling regimes.  Experimental verification is presented for the theoretical predictions.  Buckling provides an alternate route to relieve the stress built up during inflation.  Hence, buckling is a protective fail safe mechanism and it delays rupture. Further analysis in the post buckling regime incorporating twist and experiments on rupture after buckling can be pursued in the future. While the present work has been limited to isotropic elasticity, work is in progress  on anisotropic reinforced layered tubes with branches to make stronger connections with the biological problem of aneurysms.
\section*{Acknowledgement}
The  authors would like to acknowledge the support from the Natural Sciences and Engineering Research Council of Canada (NSERC) through its Discovery Grants Program.

~\newpage

\bibliography{mhsparch} 

~\newpage
\section*{Appendix A: Experimental determination of constitutive model parameters}
The parameters $\alpha_i$ and $s_i$ in  Ogden's SEF  defined in~\req{osef}, are identified from a uniaxial tensile test conducted using Instron (Model 5965). The procedure followed is according to~\cite{ogden2004}. For the uniaxial tensile test we have $\lambda_{1}=\lambda$ and from  incompressibility we have $\lambda_{2}=\lambda_{3}=\lambda^{-\frac{1}{2}}$. The nominal stress is given by
\begin{equation}
S=\frac{\partial\hat{W}(\lambda,\lambda^{-\frac{1}{2}})}{\partial \lambda}.
\end{equation}
\begin{figure}[!h]
	\center
	\includegraphics[width=0.8\textwidth]{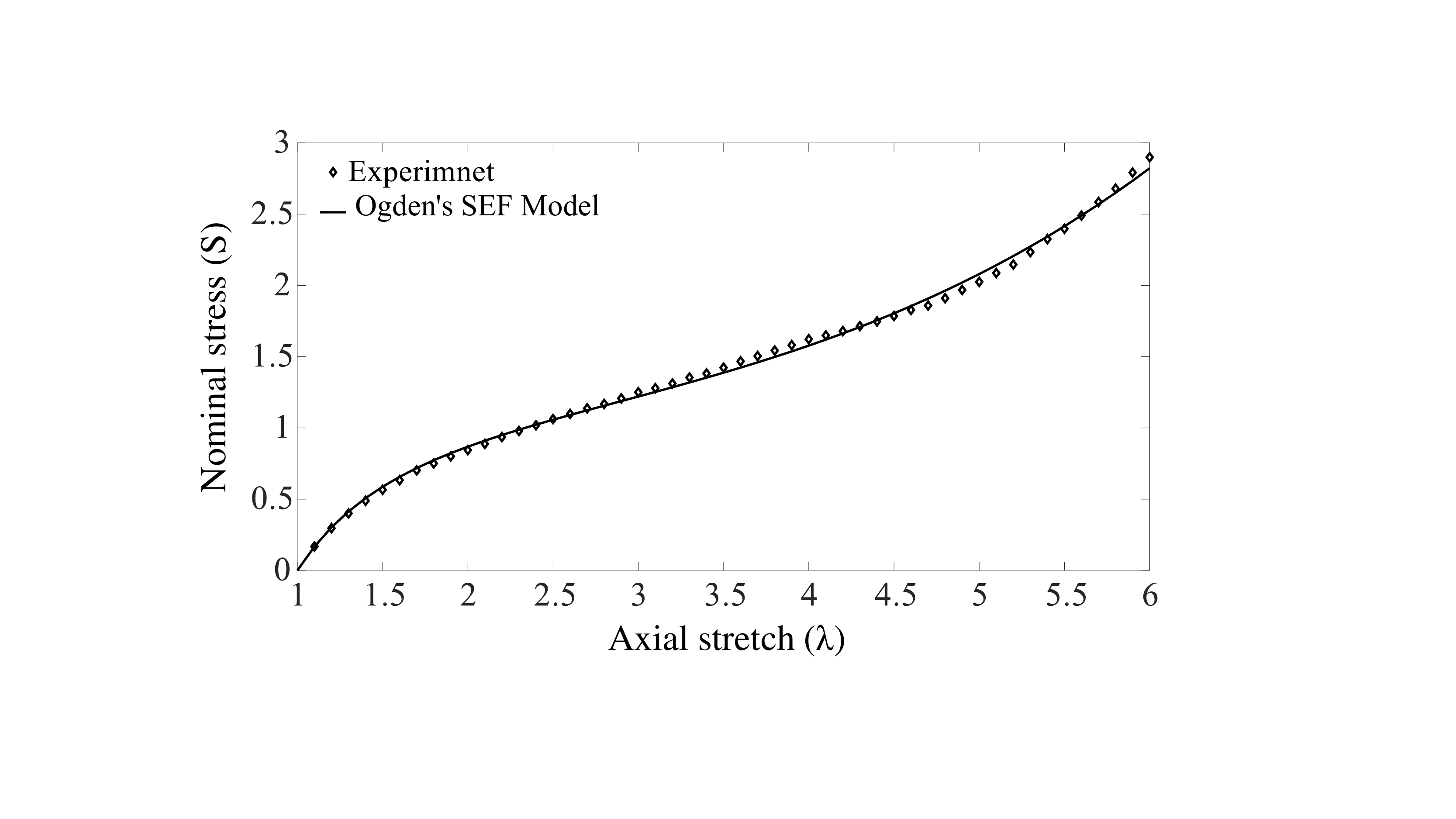}\\
	\caption{Uniaxial tensile test data compared with calibrated Ogden's SEF in~\req{osef}.}
	\label{Fig:calib}
\end{figure}
The experimental data is fitted to~\req{osef} as shown in~\fref{Fig:calib} and the identified parameters are given in Table.~1.

~\newpage
	\section*{Appendix B: Elements of the state space matrix for buckling analysis}
	{The non zero elements of $\bm{M}(\rho)$ in \req{30} are given in Table. 2 following~\cite{Chen2017}, where $\alpha=n\pi a/L$, $\Gamma$ is the Lagrange's multiplier associated with inncompressiblity, $p_0$ is a pressure like quantity from \req{assumedU}, and $B_{piqj}$ can be obtained from \req{23}. Note that $m$ and $n$ are the circumferential and axial mode numbers associated with \req{assumedU}. To obtain elements of $\bm{M}(\kappa)$ in \req{31}, we  use $\rho=\exp(\kappa)$ to affect the change variable. }
	\begin{table}[!h]
		\caption{Non-zero elements of ${\bm{M}}(\rho)$.}\label{elm}
		\begin{tabular}{p{4cm}p{4cm}p{4cm}p{4cm}}
			 ine
			$M_{11}=-1$ & $M_{12}=-m$ & $M_{13}=-\alpha\rho$\\
			 ine
			\\
			
			$M_{21}={\frac{(B_{1221}+\Gamma)m}{B_{1212}}}$ &$M_{22}={\frac{B_{1221}+\Gamma}{B_{1212}}}$ & $M_{25}={\frac{P_{0}}{B_{1212}}}$ & \\
			 ine
			$M_{31}={\frac{(B_{1331}+\Gamma)m}{B_{1313}}}\alpha\rho$ & $M_{36}={\frac{p_{0}}{B_{1313}}}$ \\
			 ine
			\\
			\multicolumn{4}{l}{$M_{41}={\frac{B_{1111}+B_{2222}-2B_{1122}+2\Gamma}{p_{0}}}+\left[{\frac{B_{2121}}{p_{0}}}-{\frac{(B_{1221}+\Gamma)^{2}}{B_{1212}p_{0}}}\right]m^{2}+\left[{\frac{B_{3131}}{p_{0}}}-{\frac{(B_{1331}+\Gamma)^{2}}{B_{1313}p_{0}}}\right]\alpha^{2}\rho^{2}$} \\
			\multicolumn{4}{l}{$M_{42}={\left[\frac{{B}_{1111}+{B}_{2121}+{B}_{2222}-2{B}_{1122}+2\Gamma}{p_{0}}-\frac{\left({B}_{1221}+p\right)^{2}}{{B}_{1212} p_{0}}\right]m}$}\\
			
			\multicolumn{2}{l}{$M_{43}={\frac{\left(B_{1111}+B_{2233}-B_{1122}-B_{1133}+\Gamma\right)\alpha\rho}{p_{0}}}$} & \\
			$M_{44}=1$ & ${M_{45}=-\frac{\left({B}_{1221}+\Gamma\right)m}{B_{1212}}}$ & {$M_{46}={-\frac{\left({B}_{1331}+\Gamma\right)\alpha\rho}{B_{1313}}}$}\\
			 ine
			\\
			
			\multicolumn{4}{l}{$M_{51}={\left[\frac{\left(B_{1111}+B_{2121}+B_{2222}-2B_{1122}+2\Gamma\right)}{p_{0}}-\frac{\left(B_{1221}+\Gamma\right)^{2}}{B_{1212} p_{0}}\right]m}$}\\
			\multicolumn{4}{l}{$M_{52}={\frac{B_{2121}}{p_{0}}-\frac{\left(B_{1221}+p\right)^{2}}{B_{1212} p_{0}}+\frac{\left(B_{1111}+B_{2222}-2B_{1122}+2 p\right)m^{2}}{p_{0}}+\frac{B_{3232}}{p_{0}} \alpha^{2}\rho^{2}}$}\\
			\multicolumn{2}{l}{$M_{53}={\frac{\left(B_{1111}+B_{2233}+B_{2332}-B_{1122}-B_{1133}+2\Gamma\right)}{p_{0}} m \alpha\rho}$} & $M_{54}=m$ & \multicolumn{1}{l}{$M_{55}=-{\frac{\left(B_{1221}+\Gamma\right)}{B_{1212}}}$}\\
			 ine\\
			\multicolumn{2}{l}{$M_{61}=-{\frac{\left(B_{1111}+B_{2233}-B_{1122}-B_{1133}+\Gamma\right)}{p_{0}} \alpha\rho}$}& \multicolumn{2}{l}{$M_{62}={\frac{\left(B_{1111}+B_{2233}+B_{2332}-B_{1122}-B_{1133}+2\Gamma\right)}{p_{0}}m\alpha\rho}$} \\
			\multicolumn{2}{l}{$M_{63}={\frac{B_{2323}}{p_{0}}m^{2}+\frac{\left(B_{1111}+B_{3333}-2B_{1133}+2\Gamma\right)}{p_{0}} \alpha^{2} \rho^{2}}$} & $M_{64}=\alpha \rho$
			%\bottomrule
		\end{tabular}
	\end{table}
	
	{Non-zero elements of $\bm{C}$ in \req{lbc} are given in Table. 3, where $P$ is the internal pressure, and $\bm{D}=\Pi^{1}_{j=N}exp({\bm{M}(\kappa_{j})}t)$ in \req{23}. Note that $\kappa_j=0$ corresponds to the inner surface $r=r_i$, and $\kappa_0=Nt$ ($N$ is the number of wall layers of thickness $t$) corresponds to the outer surface $r=r_o$.}
	
	\begin{table}[!h]
		\centering
		\caption{Non-zero elements of $\bm{C}.$} \label{celm}
		%\begin{tabular}{p{5.7cm}p{4.7cm}p{3.1cm}p{3cm}}
		\begin{tabular}{p{5.7cm}p{5.7cm}}
			$C_{i1}=D_{j1}-\frac{P}{p_{0}}\left(D_{j4}+D_{j5} m+D_{j6}\alpha\right)$ & $C_{i2}=D_{j2}-\frac{P}{p_{0}}\left(D_{j4}m+D_{j5}\right)$ \\ $C_{i3}=D_{j3}-\frac{P}{p_{0}}D_{j4}\alpha$ & $i=j-3,~j=4,5,6$
		\end{tabular}
	\end{table}

\end{document}